\documentclass[12pt,reqno,a4paper]{amsart}
\usepackage{amsmath,amssymb,dsfont,graphicx,cite}
\def\beq{\begin{equation}}
\def\eeq{\end{equation}}
\def\bea{\begin{eqnarray}}
\def\eea{\end{eqnarray}}

\newtheorem{prop}{Proposition}

\makeatletter
\expandafter\let\expandafter
\reset@font\csname reset@font\endcsname
\def\subeqnarray{\arraycolsep1pt
    \def\@eqnnum\stepcounter##1{\stepcounter{subequation}
        {\reset@font\rm(\theequation\alph{subequation})}}
\jot5mm     \eqnarray}

\makeatother
\newcommand{\CH}{{\mathcal H}}

\newcommand{\CD}{{\mathcal D}}

\newcommand{\CG}{{\mathcal G}}

\newcommand{\CM}{{\mathcal M}}

\newcommand{\CL}{{\mathcal L}}

\def\ga{\gamma}

\def\be{\beta}
\def\al{\alpha}
\def\ep{\varepsilon}
\def\l{\lambda}
\def\la{\lambda}

\def\sig{\sigma}

\def\til{\widetilde}
\def\hat{\widehat}
\def\endpf{\begin{flushright}$\square$\end{flushright}}
\def\alg{{\mathfrak g}}
\def\su2{{\mathfrak {su}}(2)}
\def\e3{{\mathfrak {e}}(3)}

\def\half{\frac{1}{2}}
\begin{document}
\title[An integrable discretization of the rational $\su2$ Gaudin model]
{\bf An integrable discretization of the rational $\su2$ Gaudin
model and related systems}
\author[MATTEO PETRERA and YURI B. SURIS]
{MATTEO PETRERA${}^{\dag}$ and YURI B. SURIS${}^{\diamond}$}

\thanks{${}^\dag$ \texttt{petrera@ma.tum.de}}
\thanks{${}^\diamond$ \texttt{suris@ma.tum.de}}

\maketitle

\centerline{\it Zentrum Mathematik,
Technische Universit\"at M\"unchen}
\centerline{\it Boltzmannstr. 3, D-85747 Garching bei M\"unchen, Germany}

\begin{abstract}
The first part of the present paper is devoted to a systematic
construction of continuous-time finite-dimensional integrable
systems arising from the rational $\su2$ Gaudin model through
certain contraction procedures. In the second part, we derive an
explicit integrable Poisson map discretizing a particular
Hamiltonian flow of the rational $\su2$ Gaudin model. Then, the
contraction procedures enable us to construct explicit integrable
discretizations of the continuous systems derived in the first
part of the paper.

\end{abstract}

\section{Introduction}

The models introduced in 1976 by M. Gaudin \cite{G1} and carrying
nowadays his name attracted considerable interest among
theoretical and mathematical physicists, playing a distinguished
role in the realm of integrable systems.

The Gaudin models describe completely integrable classical and
quantum long-range interacting spin chains. Originally the Gaudin
model was formulated \cite{G1} as a spin model related to the Lie
algebra $\mathfrak{sl}(2)$. Later it was realized \cite{G2,J} that
one can associate such a model with any semi-simple complex Lie
algebra $\alg$ and a solution of the corresponding classical
Yang-Baxter equation \cite{BD,S0}. Depending on the anisotropy of
interaction, one distinguishes between XXX, XXZ and XYZ models.
Corresponding Lax matrices turn out to depend on the spectral
parameter through rational, trigonometric and elliptic functions,
respectively. Both the classical and the quantum Gaudin models can
be formulated within the $r$-matrix approach \cite{RSTS}: they
admit a linear $r$-matrix structure, and can be seen as limiting
cases of the integrable Heisenberg magnets \cite{ST}, which admit
a quadratic $r$-matrix structure.

In the 80-es, the quantum rational Gaudin model was studied by
Sklyanin \cite{S1} and Jur\v{c}o \cite{J} from the point of view
of the quantum inverse scattering method. Precisely, Sklyanin
studied the $\su2$ rational Gaudin models, diagonalizing the
commuting Hamiltonians by means of separation of variables and
underlining the connection between his procedure and the
functional Bethe Ansatz. In \cite{Fr} the separation of variables
in the rational Gaudin model was interpreted as a geometric
Langlands correspondence. On the other hand, the algebraic
structure encoded in the linear $r$-matrix algebra allowed
Jur\v{c}o to use the algebraic Bethe Ansatz to simultaneously
diagonalize the set of commuting Hamiltonians in all cases when
$\alg$ is a generic classical Lie algebra. We have here to mention
also the the work of Reyman and Semenov-Tian-Shansky \cite{RSTS}.
Classical Hamiltonian systems associated with Lax matrices of the
Gaudin-type were widely studied by them in the context of a
general group-theoretic approach.

Some others relevants paper on the separability property of Gaudin
models are \cite{AHH,EFR,FM,Ge,KKM2,ST}. In particular, the
results in \cite{EFR,Fr} are based on the interpretation of
elliptic Gaudin models as conformal field theoretical models
(Wess-Zumino-Witten models). As a matter of fact, elliptic Gaudin
models played an important role in establishing the integrability
of the Seiberg-Witten theory \cite{SW} and in the study of
isomonodromic problems and Knizhnik-Zamolodchikov systems
\cite{FFR,Ne,Re}. Important recent work on (classical and quantum)
Gaudin models includes:

\begin{itemize}

\item In \cite{FM} the bi-Hamiltonian formulation of
$\mathfrak{sl}(n)$ rational Gaudin models has been discussed. A
pencil of Poisson brackets has been obtained that recursively
defines a complete set of integrals of motion, alternative to the
one associated with the standard Lax representation. The
constructed integrals coincide, in the $\mathfrak{sl}(2)$ case,
with the Hamiltonians of the bending flows in the moduli space of
polygons in the euclidean space introduced in \cite{KM}.

\item In \cite{HKR} an integrable time-discretization of $\su2$
rational Gaudin models has been proposed, based on the approach to
B\"acklund transformations for finite-dimensional integrable
systems developed by Sklyanin and Kuznetsov \cite{KS1}.

\item Integrable $q$-deformations of Gaudin models have been
considered in \cite{BMR} within the framework of coalgebras. Also
the superalgebra extensions of the Gaudin systems have been worked
out, see for instance \cite{BMc,KMa,MPRSq}.

\item The quantum eigenvalue problem for the $\mathfrak{gl}(n)$
rational Gaudin model has been studied and a construction for the
higher Hamiltonians has been proposed in \cite{Ta}.

\item Recently a certain interest in Gaudin models arose in the
theory of condensed matter physics. In fact, it has been noticed
\cite{AO,RSD} that the BCS model, describing the superconductivity
in metals, and the $\mathfrak{sl}(2)$ Gaudin models are closely
related.

\end{itemize}

Finally, we mention the so-called {\it algebraic extensions} of
Gaudin models, which has been studied in \cite{MPR,MPRS2,P} with
the help of a general and systematic reduction procedure based on
In\"on\"u-Wigner contractions. These extensions constitute also
the subject of the present paper, with a slightly different
derivation. Suitable algebraic and pole coalescence procedures
performed on the Gaudin Lax matrices with $N$ simple poles,
provide various families of integrable models whose Lax matrices
have higher order poles but share the linear $r$-matrix structure
with the ancestor models. This technique can be applied for any
simple Lie algebra $\mathfrak{g}$ and whatever the dependence
(rational, trigonometric, elliptic) on the spectral parameter be.
The models characterized by a single pole of increasing order $N$
and with $\mathfrak{g} = \mathfrak{su}(2)$, will be called here
the {\em one-body $\su2$ tower}. The base of the rational tower
(corresponding to $N=2$) is nothing but the {\em Lagrange top}, a
famous integrable system of classical mechanics. The many-body
counterpart of the Lagrange top is called a {\it Lagrange chain},
it is a homogeneous integrable chain of Lagrange tops with a
long-range interaction. On the other hand, the first element of
the elliptic one-body $\su2$ tower is a particular case of the
(three-dimensional) Clebsch system, describing the motion of a
free rigid body in an ideal incompressible fluid, see \cite{PR}.

A systematic approach to algebraic extensions of Gaudin models
appears independently in \cite{Ch} and \cite{MPR}. We remark that
in \cite{Ch} only $\mathfrak{sl}(n)$ Gaudin models are considered
and no $r$-matrix formulation is provided, as opposed to
\cite{MPR}.

The present paper is devoted to the construction of an integrable
time discretization of the rational $\su2$ Gaudin model and its
one-body and many-body extensions. The theory of integrable maps
got a boost when Veselov developed a theory of integrable
Lagrangian correspondences \cite{V1}, -- symplectic multi-valued
transformations possessing many independent integrals of motion in
involution. Since then the theory of integrable discretizations
has been substantially developed, a systematic presentation of the
state of the art is given in \cite{Su}. Let us mention main common
features of the discretizations found in the present paper:

\begin{itemize}

\item They are genuine birational maps, not just correspondences.

\item They preserve an invariant Poisson structure but deform
integrals, so that they are not B\"acklund transformations in the
strict sense. However they can be interpreted as B\"acklund
transformations for deformations of the original integrable
systems.

\end{itemize}

The paper is organized as follows. In Section \ref{sec2} we recall
the main features of the continuous-time rational $\su2$ Gaudin
model in order to give a systematic construction of
continuous-time one-body and many-body rational $\su2$ towers in
Section \ref{sec3}. Section \ref{sec4} is devoted to the explicit
integrable time discretization of the rational $\su2$ Gaudin
model. Then, in Section \ref{sec5}, suitable contraction
procedures on the discrete Gaudin model allow us to provide
integrable discrete-time versions of the whole one-body rational
$\su2$ tower and of the Lagrange chain. In this context, the main
goal is the {\it derivation} of continuous-time integrable systems
and their discretizations: we say practically nothing about {\it
solving} them. However, we always have in mind one of the
motivations of integrable discretizations, namely the possibility
of applying integrable Poisson maps for actual numerical
computations. Finally, some concluding remarks are contained in
Section \ref{sec6}.

Let us present here our main results. Our departure point is the
following Hamiltonian flow of the continuous-time rational $\su2$
Gaudin model:
 \beq \dot {\bf y}_i =
\left[\,\l_i\, {\bf{p}}  + {\textstyle \sum_{j=1}^N }{\bf{y}}_j\,,
\,{\bf{y}}_i\, \right], \qquad 1 \leq i \leq N, \label{1gintro}
\eeq
 where ${\bf y}_i \in \su2$, ${\bf p} \in \su2$ is a constant
matrix, and pairwise distinct numbers $\l_i$ are parameters of the
model. This flow admits $N$ independent integrals in involution:
 \beq
H_{k} = \langle \, {\bf{p}} , {\bf{y}}_k \, \rangle +
\sum_{\stackrel{\scriptstyle{j=1}} {j \neq k}}^N \frac{\langle \,
{\bf{y}}_k,{\bf{y}}_j \, \rangle }{\l_k - \l_j}, \qquad 1\le k\le
N, \label{jy}
 \eeq
 where $\langle \cdot, \cdot \rangle$ denotes the scalar product in
 $\su2\simeq\mathbb{R}^3$.

An integrable explicit discretization of the flow (\ref{1gintro})
is given by
 \beq
\widehat {{\bf{y}}}_i = \left( {\bf 1}+ \ep \, \l_i \, {{\bf{p}}}
\right) \left({\bf 1}+ \ep \, {\textstyle \sum_{j=1}^N }
{{\bf{y}}}_j \right) \,{{\bf{y}}}_i \, \left({\bf 1}+ \ep \,
{\textstyle \sum_{j=1}^N } {{\bf{y}}}_j \right)^{-1} \left( {\bf
1}+ \ep \, \l_i \, {{\bf{p}}} \right)^{-1}, \label{mg43intro}
\eeq
 with $1 \leq i \leq N$. Here hat denotes the shift $t\mapsto
t+\ep$ in the discrete time $\ep\mathbb Z$, where $\ep$ is a
(small) time step. The map (\ref{mg43intro}) is Poisson w.r.t. the
Lie-Poisson brackets on $\oplus^N \mathfrak{su}(2)^*$ and has $N$
independent and involutive integrals of motion assuring its
complete integrability:
$$
H_k(\ep) = \langle \,  {\bf{p}} ,  {\bf{y}}_k \, \rangle +
\sum_{\stackrel{\scriptstyle{j=1}} {j \neq k}}^N \frac{\langle \,
{\bf{y}}_k,{\bf{y}}_j \, \rangle }{\l_k - \l_j} \left( 1
+\frac{\ep^2}{4}  \l_k \, \l_j \, \langle \, {\bf{p}}  ,{\bf{p}}
\, \rangle   \right) -\frac{\ep}{2}
\sum_{\stackrel{\scriptstyle{j=1}} {j \neq k}}^N \, \langle \,
{\bf{p}}  , [\,{\bf{y}}_k ,{\bf{y}}_j ]\, \rangle.
$$
They are $O(\ep)$-deformations of the original ones, given in Eq. (\ref{jy}).

A contraction of $N$ simple poles to one pole of order $N$
provides the integrable flow of the one-body rational $\su2$
tower,
 \beq
 \dot {{\bf{z}}}_i = \left[\, {\bf{z}}_0, {\bf{z}}_i\,
\right] + \left[\,{\bf{p}},  {\bf{z}}_{i+1}\, \right], \qquad 0
\leq i \leq N-1, \label{1intro}
 \eeq
 with the convention ${\bf{z}}_N = {\bf{0}}$. Its integrals of motion,
 \beq
 H_{k}^{(N)}=
\langle \,{\bf{p}}, {\bf{z}}_{k} \, \rangle + \half
\sum_{i=0}^{k-1} \langle \, { \bf{z}}_i, {\bf{z}}_{k-i-1} \,
\rangle, \qquad 0 \leq k \leq N-1, \label{jy2}
 \eeq
 are in involution w.r.t. the Lie-Poisson structure obtained through a
(generalized) In\"on\"u-Wigner contraction of $\oplus^N
\mathfrak{su}(2)^*$, see eq. (\ref{np2}). An integrable
discretization of the flow (\ref{1intro}) is given by the
following map:
 \beq
 {\hat { {\bf{z}}}}_i = ({\bf 1}+\ep \,
{\bf{z}}_0) \, {\bf{z}}_i \, ({\bf 1}+\ep \,
{\bf{z}}_0)^{-1}+\ep[\,{\bf{p}},\hat{{\bf z}}_{i+1}\,]- 2
\sum_{j=2}^{N-i-1} \left( -\frac{\ep}{2}\right)^j {\rm
ad}_{\bf{p}}^j \, {\hat { {\bf{z}}}}_{i+j}, \quad 0 \leq i \leq
N-1.
 \label{mgintro}
\eeq
 This map is explicit (one can compute $\hat{{\bf z}}_i$ successively,
from $i=N-1$ to $i=0$), and Poisson w.r.t. the bracket
(\ref{np2}), preserving therefore the Casimir functions of this
bracket. Additionally, it has $N$ independent integrals of motion
in involution, assuring its complete integrability:
$$
{H}_k^{(N)} (\ep) = \langle \,{\bf{p}}, { {\bf{z}}}_{k} \, \rangle
+ \half \sum_{i=0}^{k-1} \langle \,  {{ \bf{z}}}_i, {
{\bf{z}}}_{k-i-1} \, \rangle + \frac{\ep}{2} \langle \, {\bf{p}} ,
[\,{{\bf{z}}}_0 ,  {{\bf{z}}}_k ]\, \rangle
+\frac{\ep^2}{8}\langle \,{\bf{p}}, {\bf{p}}  \, \rangle
\sum_{i=0}^{k-1} \langle \, { { \bf{z}}}_{i+1}, {  {\bf{z}}}_{k-i}
\, \rangle,
$$
with $0 \leq k \leq N-1$ (these integrals are
$O(\ep)$-deformations of (\ref{jy2})).

To stress the importance of the flow (\ref{1intro}), we note that
its simplest instance, corresponding to $N=2$, describes the
dynamics of the three-dimensional Lagrange top in the rest frame:
$$
\dot {{\bf{z}}}_0 = [\,{\bf{p}},{\bf{z}}_{1}\,], \qquad \dot
{{\bf{z}}}_1 = [\,{\bf{z}}_0 , {\bf{z}}_1\, ],
$$
with ${\bf{z}}_0\in \mathbb{R}^3$ being the vector of kinetic
momentum of the body, ${\bf{z}}_1\in \mathbb{R}^3$ being the
vector pointing from the fixed point to the center of mass of the
body, and ${\bf{p}}$ being the constant vector along the gravity
field. The Lagrange top is a Hamiltonian system w.r.t. the
Lie-Poisson bracket on $\e3^*$, with the Hamiltonian function
$$
H_1^{(2)} = \langle \,{\bf{p}}, {\bf{z}}_{1} \, \rangle + \half
\langle \, { \bf{z}}_0, {\bf{z}}_{0} \, \rangle.
$$
Its complete integrability is ensured by the second integral of
motion $H_0^{(2)} = \langle \,{\bf{p}}, {\bf{z}}_{0} \, \rangle$,
and by the Casimir functions $C_{0}^{(2)} = \langle \, {\bf{z}}_0,
{\bf{z}}_1 \, \rangle$ and $C_{1}^{(2)} = \half \langle \,
{\bf{z}}_1,  {\bf{z}}_1 \, \rangle$. The map (\ref{mgintro}) for
$N=2$ coincides with the integrable discretization of the Lagrange
top found in \cite{BS}:
$$
{\hat { {\bf{z}}}}_0 = {\bf{z}}_0+\ep[\,{\bf{p}},\hat{{\bf
z}}_1\,]\,,\qquad {\hat { {\bf{z}}}}_1 = ({\bf 1}+\ep \,
{\bf{z}}_0) \, {\bf{z}}_1 \, ({\bf 1}+\ep \, {\bf{z}}_0)^{-1},
$$
with the deformed Hamiltonian function
$$
{H}_1^{(2)} (\ep) = \langle \,{\bf{p}}, { {\bf{z}}}_{1} \, \rangle
+ \half \langle \,  {{ \bf{z}}}_0, { {\bf{z}}}_{0} \, \rangle +
\frac{\ep}{2} \langle \, {\bf{p}} , [\,{{\bf{z}}}_0 , {{\bf{z}}}_1
]\, \rangle,
$$
(all other integrals remain non-deformed in this case).

A contraction of $N=2M$ simple poles to $M$ double poles provides the
integrable flow of the Lagrange chain,
$$
\dot {\bf{m}}_i = \left[\,  {\bf{p}} , {\bf{a}}_i \, \right] +
\left[\, \mu_i \, {\bf{p}} +  {\textstyle \sum_{k=1}^M }
{\bf{m}}_k , {\bf{m}}_i \, \right] ,\quad \dot {\bf{a}}_i =
\left[\,\mu_i \, {\bf{p}} +  {\textstyle \sum_{k=1}^M } {\bf{m}}_k
 , {\bf{a}}_i \, \right] \, ,  \qquad 1 \leq i \leq M.
$$
Here $({\bf{m}}_i, {\bf{a}}_i) \in \mathfrak{e}(3)^*$ and
$\mu_i$'s are free parameters of the model. (In particular, for
$M=1$ and $\mu_1=0$, one recovers again the Lagrange top, upon the
re-naming ${\bf z}_0\mapsto{\bf m}_1$ and ${\bf z}_1\mapsto{\bf
a}_1$.) The Lagrange chain possesses $2M$ independent integrals of
motion in involution, given in Eqs. (\ref{567},\ref{568}). An
explicit discretization is given by  \bea && \! \! \! \!\!\!\!\!
\widehat {{\bf{m}}}_i = \left( {\bf 1}+ \ep \, \mu_i \, {{\bf{p}}}
\right) \left({\bf 1}+ \ep \,  {\textstyle \sum_{j=1}^M }
{{\bf{m}}}_j \right) \,{{\bf{m}}}_i \, \left({\bf 1}+ \ep \,
{\textstyle \sum_{j=1}^M } {{\bf{m}}}_j \right)^{-1} \left( {\bf
1}+ \ep \, \mu_i \, {{\bf{p}}} \right)^{-1} + \ep \, \left[\,
{{\bf{p}}},
\widehat {{\bf{a}}}_i\, \, \right], \nonumber \\
&& \! \! \! \!\!\!\!  \widehat {{\bf{a}}}_i = \left( {\bf 1}+ \ep
\, \mu_i \, {{\bf{p}}} \right) \left({\bf 1}+ \ep \, {\textstyle
\sum_{j=1}^M }  {{\bf{m}}}_j \right) \,{{\bf{a}}}_i \, \left({\bf
1}+ \ep \, {\textstyle \sum_{j=1}^M }  {{\bf{m}}}_j \right)^{-1}
\left( {\bf 1}+ \ep \, \mu_i \, {{\bf{p}}} \right)^{-1} ,
\nonumber \eea with $1 \leq i \leq M$. Expressions for the
integrals of motion of this Poisson map are given in Eqs.
(\ref{567dd},\ref{568dd}), they are $O(\ep)$-deformations of the
integrals of the continuous system.

\section{The continuous-time rational $\su2$ Gaudin model} \label{sec2}

The aim of this Section is to give a terse survey of the main
features of the continuous-time rational $\su2$ Gaudin model. In
particular, we give its Lax representation along with the
interpretation of the latter in terms of the (linear) $r$-matrix
structure. For further details we refer to \cite{G1,G2,J,RSTS}.

Let us choose the following basis of the linear space $\su2$:
$$
\sigma_1 =  \half \left(\begin{array}{cc}
0 & -{\rm{i}}  \\
-{\rm{i}} & 0
\end{array}\right), \qquad
\sigma_2 =  \half \left(\begin{array}{cc}
0 & -1  \\
1 & 0
\end{array}\right), \qquad
\sigma_3 = \half \left(\begin{array}{cc}
-{\rm{i}} & 0  \\
0 & {\rm{i}}
\end{array}\right).
$$
We recall that the correspondence
$$
\mathbb{R}^3 \ni {\bf{a}} = (a^1,a^2,a^3) \; \longleftrightarrow \;
{\bf{a}} = \half \left(\begin{array}{cc}
-{\rm{i}} \,a^{3} &  -{\rm{i}} \,a^{1}-a^2 \\
-{\rm{i}} \,a^{1}+ a^2 & {\rm{i}} \, a^{3}
\end{array}\right)=a^\alpha\sigma_\alpha \, \in \mathfrak{su}(2),
$$
is an isomorphism between $(\mathfrak{su}(2),[\,\cdot, \cdot\,])$
and the Lie algebra $(\mathbb{R}^3, \times)$, where $\times$
stands for the vector product. (Here and below we assume the
summation over the repeated Greek indices.) This allows us to
identify vectors from $\mathbb{R}^3$ with matrices from $\su2$. We
supply $\mathfrak{su}(2)$ with the scalar product $\langle \,
\cdot, \cdot \, \rangle$ induced from $\mathbb{R}^3$, namely $
\langle \, {\bf{a}}, {\bf{b}}  \, \rangle = -2 \, {\rm{tr}} \,
({\bf{a}} {\bf{b}}) = 2 \, {\rm{tr}} \, ( {\bf{b}}
{\bf{a}}^\dag), \; \forall \, {\bf{a}}, {\bf{b}}  \in \su2$. The
matrix multiplication and the commutator in $\su2$ are related by
the following formula:
\begin{equation} \label{lemm1}
{\bf{a}} \, {\bf{b}} = -\frac14 \langle \, {\bf{a}}, {\bf{b}} \,
\rangle {\bf 1} + \half [\, {\bf{a}} , {\bf{b}} \,], \qquad
\forall{\bf{a}},{\bf{b}} \in \su2.
\end{equation}
In particular, if $\langle \, {\bf{a}}, {\bf{b}} \, \rangle=0$,
then ${\bf{a}} {\bf{b}} + {\bf{b}} {\bf{a}} =0$.

The above scalar product allows us to identify the dual space
$\mathfrak{su}(2)^*$ with $\su2$, so that the coadjoint action of
the algebra becomes the usual Lie bracket with minus, i.e. $ {\rm
ad}_{\bf{b}}^* \, {\bf{a}} =[\,{\bf{a}},{\bf{b}}\,]= - {\rm
ad}_{\bf{b}} \, {\bf{a}}$, with ${\bf{a}},{\bf{b}} \in \su2$.
\smallskip

We will denote by $\{y^\al_i \}_{\al=1}^{3}$, $1 \leq i \leq N$,
the coordinate functions (in the basis $\sigma_\alpha$) on the
$i$-th copy of $\mathfrak{su}(2)^*$ in $\oplus^N
\mathfrak{su}(2)^*$. So, ${\bf{y}}_i  =y_i^\al\sig_\al\,$. In
these coordinates, the Lie-Poisson bracket on  $\oplus^N
\mathfrak{su}(2)^*$ reads
\beq
 \left\{ y^\al_i, y^\be_j \right\}=
- \delta_{i,j} \, \epsilon_{\al \be \gamma} \, {y}^\gamma_{i},
\label{LPgdd}
 \eeq
with $1\leq i,j \leq N$. Here  $\delta_{i,j}$ is the standard
Kronecker symbol and $\epsilon_{\al \be \gamma}$ is the
skew-symmetric tensor with $\epsilon_{1 2 3}=1$. The bracket
(\ref{LPgdd}) possesses  $N$ Casimir functions \beq C_{i} = \half
\langle \, {\bf{y}}_i,{\bf{y}}_i \, \rangle, \qquad 1 \leq i \leq
N. \label{poi} \eeq Fixing their values, we get a symplectic leaf
where the Lie-Poisson bracket is non-degenerate. It is a union of
$N$ two-dimensional spheres.

The continuous-time rational $\su2$ Gaudin model is governed by
the following rational Lax matrix from the loop algebra $\su2
[\,\l,\l^{-1}]$: \beq \CL_{\CG}(\l)  = {\bf{p}} + \sum_{i=1}^{N}
\frac{{\bf{y}}_i }{\l - \l_i}\, , \label{lgaure} \eeq where the
$\lambda_i$'s, with $\lambda_i \neq \lambda_k, 1 \leq i,k \leq N$,
are complex parameters of the model, and ${\bf{p}}\in\su2$ is a
constant vector. This Lax matrix yields a completely integrable
system on the Lie-Poisson manifold $\oplus^N \mathfrak{su}(2)^*$.
In particular, its spectral invariants are in involution. This can
be demonstrated with the help of a linear $r$-matrix formulation.
We quote the following result \cite{J}.

\begin{prop} The Lax matrix (\ref{lgaure})
satisfies the linear $r$-matrix relation \beq
 \left\{ \CL_\CG(\la)
\otimes {\bf 1}, {\bf 1} \otimes \CL_\CG(\mu)
\right\}+\left[\,r(\la-\mu), \CL_\CG(\la) \otimes {\bf 1}+{\bf 1}
\otimes \CL_\CG(\mu)\, \right]=0, \quad \forall \, \l, \mu \in
\mathbb{C}, \label{pb4}
 \eeq
with
 \beq r(\l) = -   \frac{1}{\l} \sigma_\al \otimes \sigma_\al.
\label{pb43ui}
 \eeq
The $r$-matrix (\ref{pb43ui}) is equivalent to $r(\l)=-\Pi / (2 \,
\l)$, where $\Pi$ is the permutation operator in $\mathbb{C}^2
\otimes \mathbb{C}^2$.
\end{prop}

The spectral invariants of $ \CL_{\CG}(\l)$ are the coefficients
of its characteristic equation $\det( \CL_{\CG}(\l) -\mu \, {\bf
1})=0$, which reads
$$
-\mu^2 = \frac14 \langle{\bf{p}}, {\bf{p}} \rangle + \half
\sum_{i=1}^N \left[\frac{H_i}{\l-\l_i}+
\frac{C_i}{(\l-\l_i)^2}\right].
$$
Here $C_i$ are the Casimir functions given in Eq. (\ref{poi}),
whereas the functions
 \beq
 H_{i} = \langle \,
{\bf{p}} , {\bf{y}}_i \, \rangle +
\sum_{\stackrel{\scriptstyle{j=1}} {j \neq i}}^N \frac{\langle \,
{\bf{y}}_i,{\bf{y}}_j \, \rangle }{\l_i - \l_j}, \qquad 1\le i\le
N,  \label{35}
 \eeq
are the independent and involutive Hamiltonians of the rational
$\su2$ Gaudin model. We shall focus our attention on Hamiltonians
obtained as linear combinations of the integrals $H_{i}$:
 \beq
\sum_{i=1}^N \eta_i \, H_i=\half
\sum_{\stackrel{\scriptstyle{i,j=1}} {i \neq j}}^N \frac{\eta_i -
\eta_j}{\l_i - \l_j} \, \langle \,  {\bf{y}}_i , {\bf{y}}_j \,
\rangle + \sum_{i=1}^N \eta_i \, \langle \,  {\bf{p}} , {\bf{y}}_i
\, \rangle.\label{itglrrrr}
 \eeq
An important specialization of the Hamiltonian (\ref{itglrrrr}) is
obtained considering $\eta_i= \l_i$, $1 \leq i \leq N$. It reads
\beq
 \CH_\CG = \half \sum_{\stackrel{\scriptstyle{i,j=1}} {i \neq
j}}^N \langle \,  {\bf{y}}_i , {\bf{y}}_j \, \rangle +
\sum_{i=1}^N \l_i \, \langle \,  {\bf{p}} , {\bf{y}}_i \,
\rangle.\label{itglt}
 \eeq
From the physical point of view it describes an interaction of
$\su2$ vectors ${\bf{y}}_i$ (spins in the quantum case) with a
homogeneous and constant external field ${\bf{p}}$. One verifies
by a direct computation that the Hamiltonian flow generated by the
integral (\ref{itglt}) is given by \beq
 \dot {\bf y}_i =
\left[\,\l_i\, {\bf{p}}  + {\textstyle \sum_{j=1}^N }{\bf{y}}_j\,,
\,{\bf{y}}_i\, \right], \qquad 1 \leq i \leq N. \label{1g}
 \eeq
Eq. (\ref{1g}) admits the following Lax representation:
\beq
 \dot
\CL_{\CG} (\l) =\left[\,  \CL_{\CG}(\l), \CM_{\CG}^{(-)} (\l) \,
\right]=-\left[\,  \CL_{\CG}(\l), \CM_{\CG}^{(+)} (\l)  \,
\right], \label{kh}
 \eeq
with the matrix $\CL_{\CG}(\l)$ given in Eq. (\ref{lgaure}) and
\beq \CM_{\CG}^{(-)} (\l) = \sum_{i=1}^{N} \frac{\l_i \,
{\bf{y}}_i}{\l - \l_i}, \qquad \CM_{\CG}^{(+)} (\l) = \l  \,
{\bf{p}} + \sum_{i=1}^{N}{\bf{y}}_i . \label{100g} \eeq

\section{Contractions of rational $\su2$ Gaudin models}
\label{sec3}

\subsection{Contraction of the Lie-Poisson algebra $\oplus^{N}\mathfrak{su}(2)^*$}
\label{p5}

The following statement allows one to get the generalized
In\"on\"u-Wigner contraction of the direct sum of $N$ copies of
$\mathfrak{su}(2)^*$ \cite{IW,MPR,eve}. It shall enable us to
construct the {\it{rational one-body $\su2$ tower}} in Subsection
\ref{df3}. See also \cite{MPR,MPRS2} for further details.

\begin{prop}\label{758}
Consider the Lie-Poisson bracket (\ref{LPgdd}) of $\oplus^N
\mathfrak{su}(2)^*\simeq(\mathbb{R}^3)^N$ with coordinates $({\bf
y}_j)_{j=1}^N$, and a linear map $(\mathbb{R}^3)^N \to
(\mathbb{R}^3)^N$, $({\bf y}_j)_{j=1}^N\mapsto ({\bf
z}_i)_{i=0}^{N-1}$, given by \beq {\bf z}_i= \vartheta^i \
\sum_{j=1}^N \nu_j^i \, {\bf y}_j, \qquad 0 \leq i \leq N-1,
\label{ty2} \eeq with pairwise distinct $\nu_j \in \mathbb{C}$ and
$0 < \vartheta \leq 1$ {\rm{(contraction parameter)}}. Then the
bracket induced on $(\mathbb{R}^3)^N$ with coordinates $({\bf
z}_i)_{i=0}^{N-1}$ under the map (\ref{ty2}) is regular for
$\vartheta\to 0$, and tends in this limit to \beq \left\{z^\al_i,
z^\be_j \right\}= \left\{
\begin{array}{cc}
-\epsilon_{\al \be \ga}\, z^\gamma_{i+j} & \quad i+j < N,\\
0 & \quad i+j \geq N,
\end{array} \right.\label{np2}
\eeq with $0\leq i,j \leq N-1$. We shall denote the Lie-Poisson
algebra (\ref{np2}) by $\mathcal{C}_N(\mathfrak{su}(2)^*)$.
\end{prop}
{\bf{Proof:}} Using Eqs. (\ref{LPgdd}) and (\ref{ty2}) we get:
\bea \left\{\, z^\al_i, z^\be_j \, \right\}_\vartheta &=&
\vartheta^{i+j} \, \sum_{n,m=1}^N  \nu_n^i \, \nu_m^j
\left\{\, y^\al_n,  y^\be_m \, \right\}= \nonumber \\
&=& -\epsilon_{\al \be \gamma} \, \vartheta^{i+j} \,
\sum_{n=1}^N  \nu_n^{i+j} \  y^\ga_n = \left\{ \begin{array}{cc}
-\epsilon_{\al \be \ga} \, z^\gamma_{i+j} & \quad i+j < N,\\
O(\vartheta) & \quad i+j \geq N.\nonumber
\end{array} \right.
\eea The limit $\vartheta \rightarrow 0$ leads to (\ref{np2}). It
is easy to check that the antisymmetric bracket  (\ref{np2})
satisfies the Jacobi identity.
\endpf

The following $N$ functions are Casimirs for the Lie-Poisson
bracket (\ref{np2}): \beq C_{k}^{(N)}= \half  \sum_{i=k}^{N-1}
\langle \, {\bf {z}}_i , {\bf z}_{N+k-i-1} \, \rangle,\qquad 0
\leq k \leq N-1.\label{LP75} \eeq

We illustrate this construction by the cases of small $N$. For
$N=2$ the contracted bracket $\mathcal{C}_2(\mathfrak{su}(2)^*)$
reads \beq
 \left\{ z_0^{\alpha}, z_0^{\beta}   \right\}= -\ep_{\al \be \gamma} \,
z_0^{\gamma}, \qquad \left\{    z_0^{\alpha}, z_1^{\beta}
\right\}= -\ep_{\al \be \gamma} \, z_1^{\gamma}, \qquad \left\{
z_1^{\alpha}, z_1^{\beta}   \right\}= 0. \label{E3}
 \eeq
This is the Lie-Poisson bracket of $\e3^*=\su2^* \oplus_s
\mathbb{R}^3$. Its Casimir functions are \beq C_{0}^{(2)} =
\langle \, {\bf{z}}_0,  {\bf{z}}_1 \, \rangle, \qquad C_{1}^{(2)}
= \half \langle \,  {\bf{z}}_1,  {\bf{z}}_1 \, \rangle. \label{67}
\eeq
 For $N=3$ we get the contracted Lie-Poisson bracket
$\mathcal{C}_3(\mathfrak{su}(2)^*)$:
\begin{subequations}
\begin{align}
& \left\{    z_0^{\alpha}, z_0^{\beta}   \right\}
= -\ep_{\al \be \gamma} \, z_0^{\gamma}, \qquad
\left\{    z_0^{\alpha}, z_1^{\beta}   \right\}
= -\ep_{\al \be \gamma} \, z_1^{\gamma}, \qquad
\left\{    z_0^{\alpha}, z_2^{\beta}   \right\}
= -\ep_{\al \be \gamma} \, z_2^{\gamma},
\label{e3p1} \\
& \qquad \qquad \left\{    z_1^{\alpha}, z_1^{\beta}   \right\}
=-\ep_{\al \be \gamma} \,  z_2^{\gamma}, \qquad
\left\{    z_1^{\alpha}, z_2^{\beta}   \right\}= 0, \qquad
\left\{    z_2^{\al}, z_2^{\be}   \right\}= 0. \label{e3p2}
\end{align}
\end{subequations}
Its Casimir functions are \beq C_{0}^{(3)} = \langle \,
{\bf{z}}_0, {\bf{z}}_2 \, \rangle + \half \langle \,  {\bf{z}}_1,
{\bf{z}}_1 \, \rangle, \qquad
  C_{1}^{(3)} = \langle \,  {\bf{z}}_1,  {\bf{z}}_2 \, \rangle, \qquad
 C_{2}^{(3)} = \half \langle \,  {\bf{z}}_2, {\bf{z}}_2 \, \rangle. \nonumber
 \eeq

The following result will be useful in the next Sections.

\begin{prop} \label{invol}
Let $H,G$ be two involutive functions w.r.t. the Lie-Poisson
brackets (\ref{LPgdd}) on $\oplus^N \mathfrak{su}(2)^*$. If $\til
H,\til G$ are the corresponding functions on
$\mathcal{C}_N(\mathfrak{su}(2)^*))$ obtained from $H,G$ by
applying the map (\ref{ty2}) in the contraction limit $\vartheta
\rightarrow 0$, then they are in involution w.r.t. the Lie-Poisson
brackets (\ref{np2}).
\end{prop}

{\bf{Proof:}} In the local coordinates $\{y^\al_i \}_{\al=1}^{3}$, $1 \leq i \leq N$,
we have:
\bea
0=\left\{H, G \right\}&=& \sum_{i,j=1}^N
\frac{\partial H}{\partial y^\al_i } \, \frac{\partial G}{\partial y^\be_j }\,
\left\{y^\al_i, y^\be_j \right\}=
-\epsilon_{\al \be \ga} \, \sum_{i=1}^N
\frac{\partial H}{\partial y^\al_i } \,
\frac{\partial G}{\partial y^\be_i }\,y^\ga_i = \nonumber \\
&=&
-\epsilon_{\al \be \ga} \, \sum_{i=1}^N \sum_{n,m=0}^{N-1}
\frac{\partial \til H}{\partial z^\al_n } \, \frac{\partial \til G}{\partial z^\be_m }\,
\vartheta^{n+m} \, \nu^{n+m}_i\,y^\ga_i = \nonumber \\
&=&
 -\epsilon_{\al \be \ga} \, \sum_{\stackrel{\scriptstyle{n,m=0}} {n+m < N}}^{N-1}
\frac{\partial \til H}{\partial z^\al_n } \, \frac{\partial \til
G}{\partial z^\be_m }\, z^\ga_{n+m} + O(\vartheta), \nonumber \eea
where the first term does not depend explicitly on the contraction
parameter $\vartheta$. Performing the limit $\vartheta \rightarrow
0$ we get $\{\til H, \til G \}=0$.

\endpf

\subsection{Contraction of the Lie-Poisson algebra $\oplus^{NM}\mathfrak{su}(2)^*$}
\label{p6}

The following Proposition enables one to get a Lie-Poisson algebra
given by the direct sum of $M$ copies of
$\mathcal{C}_N(\mathfrak{su}(2)^*)$ directly from the Lie-Poisson
algebra $\oplus^{NM}\mathfrak{su}(2)^*$ associated with a
$NM$-body Gaudin model. Its specialization to $M=1$ is equivalent
to Proposition \ref{758}.

\begin{prop} \label{ji}
Consider the Lie-Poisson brackets of $\oplus^{NM}
\mathfrak{su}(2)^*\simeq(\mathbb{R}^3)^{NM}$ with the coordinates
$({\bf y}_j)_{j=1}^{NM}$, and a linear map $(\mathbb{R}^3)^{NM}
\rightarrow (\mathbb{R}^3)^{NM}$, $({\bf y}_j)\mapsto ({\bf
z}_{i,n})$, given by \beq
 {\bf z}_{i,n}= \vartheta^i \,  \sum_{j=1}^N  \nu_{N(n-1)+j}^i \, {\bf y}_{N(n-1)+j},
\qquad  1 \leq n \leq M, \quad 0 \leq i \leq N-1, \label{ty234}
\eeq with pairwise distinct $\nu_j \in \mathbb{C}$ and $0 <
\vartheta \leq 1$. Then the bracket induced on
$(\mathbb{R}^3)^{NM}$ with coordinates $({\bf z}_{i,n})$ under the
map (\ref{ty234}) is regular for $\vartheta\to 0$, and tends in
this limit $\vartheta \rightarrow 0$ to \beq \left\{z^\al_{i,n},
z^\be_{j,m} \right\}= \left\{
\begin{array}{cc}
-\delta_{n,m} \, \epsilon_{\al \be \ga} \, z^\gamma_{i+j,n} & \quad i+j < N,\\
0 & \quad i+j \geq N,
\end{array} \right.\label{tyR2}
\eeq with $0\leq i,j \leq N-1$ and $1 \leq n,m \leq M$. We shall
denote the Lie-Poisson algebra (\ref{tyR2}) by
$\oplus^M\mathcal{C}_N(\mathfrak{su}(2)^*)$.
\end{prop}

{\bf{Proof:}} Using Eqs. (\ref{LPgdd}) and (\ref{ty234}) we get:
\bea \left\{z^\al_{i,n}, z^\be_{j,m} \right\}_\vartheta &=&
\vartheta^{i+j}  \sum_{l,k=1}^N   \nu_{N(n-1)+l}^i \,
\nu_{N(m-1)+k}^j \,
\left\{ y^\al_{N(n-1)+l}, y^\be_{N(m-1)+k} \right\}= \nonumber \\
&=&- \epsilon_{\al \be \ga}  \, \vartheta^{i+j}  \sum_{l,k=1}^N
\nu_{N(n-1)+l}^i \, \nu_{N(m-1)+k}^j \, \delta_{n,m}
\delta_{l,k}\, y^\ga_{N(n-1)+l}. \nonumber \\ &=&- \delta_{n,m} \,
\epsilon_{\al \be \ga} \,\vartheta^{i+j} \sum_{l=1}^N
\nu_{N(n-1)+l}^{i+j}\, y^\ga_{N(n-1)+l} = \nonumber \\
 &=&\left\{ \begin{array}{cc}
-\delta_{n,m} \, \epsilon_{\al \be \ga} \, z^\gamma_{i+j,n} & \quad i+j < N,\\
O(\vartheta) & \quad i+j \geq N.
\end{array} \right. \nonumber
\eea
The limit $\vartheta \rightarrow 0$ leads to (\ref{tyR2}).
\endpf

The Lie-Poisson brackets (\ref{tyR2}) have $NM$ Casimir functions
of the form (\ref{LP75}).

A computation similar to the one in the proof of Proposition
\ref{invol} leads to the following statement.

\begin{prop} \label{7584t}
Let $H,G$ be two involutive functions w.r.t. the Lie-Poisson
brackets (\ref{LPgdd}) on $\oplus^{NM} \mathfrak{su}(2)^*$. If
$\til H,\til G$ are the corresponding functions on $\oplus^{M}
\mathcal{C}_N(\mathfrak{su}(2)^*)$ obtained from $H,G$ by applying
the map (\ref{ty234}) in the contraction limit $\vartheta
\rightarrow 0$, then they are in involution w.r.t. the Lie-Poisson
bracket (\ref{tyR2}).
\end{prop}

\subsection{The rational one-body $\su2$ tower} \label{df3}

Our aim is now to apply the map (\ref{ty2}), in the contraction
limit $\vartheta \rightarrow 0$, to the Lax matrix (\ref{lgaure}),
in order to get a new rational Lax matrix governing the rational
{\it {one-body $\su2$ tower}}. To do so a second ingredient is
needed: as shown in \cite{KPR,MPR} we have to consider the {\it
pole coalescence} $\l_i =\vartheta \, \nu_i$, $1 \leq i \leq N$.
This pole fusion can be considered as the analytical counterpart
of the algebraic one given by the map (\ref{ty2}).

\begin{prop} \label{lo}
Consider the Lax matrix (\ref{lgaure}) with $\l_i = \vartheta \,
\nu_i , \;1 \leq i \leq N$. Under the map (\ref{ty2}) and upon the
limit $\vartheta \rightarrow 0$ the Lax matrix (\ref{lgaure})
tends to \beq \CL_N (\l) = {\bf p}+
 \sum_{i=0}^{N-1} \frac{ {\bf z}_i}{\l^{i+1}},  \label{LP7}
\eeq while the Lax equation (\ref{kh}) turns into \beq \dot
\CL_{N}  (\l) =\left[\,  \CL_{N}(\l), \CM_{N}^{(-)}(\l) \,
\right]= - \left[\,  \CL_{N} (\l), \CM_{N}^{(+)} (\l)  \, \right],
\label{10mm} \eeq with \beq \CM_{N}^{(-)}  (\l) = \sum_{i=1}^{N-1
} \frac{  {\bf{z}}_i}{\l^i}, \qquad \CM_{N}^{(+)}  (\l) = \l \,
{\bf{p}} + {\bf{z}}_0. \nonumber \eeq The Lax matrix (\ref{LP7})
satisfies the linear $r$-matrix relation (\ref{pb4}) with the same
$r$-matrix (\ref{pb43ui}).
\end{prop}

{\bf{Proof:}}  The first part of Proposition \ref{lo} can be
proved by applying the map  (\ref{ty2}) and the pole coalescence
$\l_i = \vartheta \, \nu_i , \;1 \leq i \leq N$, on Eqs.
(\ref{lgaure}) and (\ref{100g}). We get
$$
\CL_{\CG}(\l) =
{\bf{p}} + \sum_{j=1}^{N} \frac{{\bf{y}}_j }{\l - \vartheta \, \nu_j}=
\frac{1}{\l}\sum_{j=1}^{N} \sum_{i=0}^{N-1}  \left(\frac{\vartheta \,
\nu_j}{\l}\right)^{i} {\bf{y}}_j + O(\vartheta)
\xrightarrow{\scriptstyle {\vartheta \rightarrow 0}} \CL_N (\l),
$$
and
\bea
&&\CM_{\CG}^{(-)}(\l) =\sum_{j=1}^{N} \frac{\vartheta \, \nu_j \,
{\bf{y}}_j}{\l - \vartheta \, \nu_j}=
\sum_{j=1}^{N} \sum_{i=0}^{N-2}  \left(\frac{\vartheta \, \nu_j}{\l}\right)^{i+1}
{\bf{y}}_j + O(\vartheta)
\xrightarrow{\scriptstyle {\vartheta \rightarrow 0}}
 \sum_{i=0}^{N-2} \frac{{\bf{z}}_{i+1}}{\l^{i+1}}= \CM_{N}^{(-)}  (\l), \nonumber \\
&&\CM_{\CG}^{(+)}(\l)=\l  \, {\bf{p}} + \sum_{i=1}^{N}{\bf{y}}_i=\CM_{N}^{(+)}  (\l).
\nonumber
 \eea

The fact that the Lax matrix (\ref{LP7}) satisfies the linear
$r$-matrix relation (\ref{pb4}) with the same $r$-matrix
(\ref{pb43ui}) requires a longer but straightforward computation.
We refer to \cite{MPR,MPRS2,P} for a detailed proof.

\endpf

The Hamiltonian flow described by the Lax equation (\ref{10mm}) is
given by \beq \dot {{\bf{z}}}_i = \left[\, {\bf{z}}_0,
{\bf{z}}_i\, \right] + \left[\,{\bf{p}},  {\bf{z}}_{i+1}\,
\right], \qquad 0 \leq i \leq N-1, \label{1} \eeq with $
{\bf{z}}_N = {\bf{0}}$, while the characteristic equation of the
Lax matrix $\det( \CL_N(\l) -\mu \, {\bf 1})=0$ reads \beq -\mu^2=
\frac{1}{4} \langle \,{\bf{p}}, {\bf{p}} \, \rangle + \half
\sum_{k=0}^{N-1} \frac{H_{k}^{(N)}}{\l^{k+1}} +\half
\sum_{k=0}^{N-1} \frac{C_{k}^{(N)} }{\l^{k+N+1}}, \nonumber \eeq
where the functions $C_{k}^{(N)}$, $0 \leq k \leq N-1$, are the
Casimir functions (\ref{LP75}), while the functions  \beq
H_{k}^{(N)}= \langle \,{\bf{p}}, {\bf{z}}_{k} \, \rangle + \half
\sum_{i=0}^{k-1} \langle \, { \bf{z}}_i, {\bf{z}}_{k-i-1} \,
\rangle,  \label{34} \eeq are the $N$ independent involutive
Hamiltonians of the rational one-body $\su2$ tower.

Notice that it is possible to obtain the integrals (\ref{34})
using the map (\ref{ty2}), in the contraction limit $\vartheta
\rightarrow 0$, and the pole coalescence $\l_i = \vartheta \,
\nu_i , \;1 \leq i \leq N$, from the integrals (\ref{35}). Let us
fix $i$ such that $0 \leq i \leq N-1$. We get \bea \sum_{k=1}^{N}
\vartheta^i \, \nu_k^i \, H_k&=& \sum_{k=1}^{N} \vartheta^i \,
\nu_k^i \langle \,{\bf{p}}, {\bf{y}}_{k} \, \rangle + \half
\sum_{\stackrel{\scriptstyle{j,k=1}} {j \neq k}}^N \vartheta^{i-1}
\frac{\nu_k^i  -  \nu_j^i } {\nu_k  - \nu_j}
\langle \,{\bf{y}}_k,   {\bf{y}}_{j} \, \rangle = \nonumber \\
&=& \sum_{k=1}^{N} \vartheta^i \, \nu_k^i \langle \,{\bf{p}}, {\bf{y}}_{k} \,
\rangle +
\half \sum_{m=0}^{i-1} \sum_{\stackrel{\scriptstyle{j,k=1}} {j \neq k}}^N
 ( \vartheta \,  \nu_k)^m ( \vartheta \, \nu_j) ^{i-m-1}
 \langle \,{\bf{y}}_k, {\bf{y}}_{j} \, \rangle = \nonumber \\
&=& \langle \,{\bf{p}},  {\bf{z}}_{i} \, \rangle + \half  \sum_{m=0}^{i-1}
\langle \, \ { \bf{z}}_m,  {\bf{z}}_{i-m-1} \, \rangle = H_{i}^{(N)}. \nonumber
\eea
In the above computation we have taken into account the polynomial identity
$$
 \nu_k^i  -  \nu_j^i= ( \nu_k  -  \nu_j)
 \sum_{m=0}^{i-1}  \nu_k^m \, \nu_j^{i-m-1}.
$$

The contracted version of the Hamiltonian (\ref{itglt}) is given
by $H_{1}^{(N)}$, namely the integral of motion generating the
Hamiltonian flow given in Eq. (\ref{1}), while the contracted
version of the linear integral $\sum_{k=1}^N  H_k  = \sum_{k=1}^N
\langle \,  {\bf{p}} , {\bf{y}}_k \, \rangle$ is given by
$H_{0}^{(N)}$.

Let us remark that the involutivity of the spectral invariants of
the Lax matrix $\CL_N(\l)$ is indeed ensured thanks to the
$r$-matrix formulation (\ref{pb4}). Their involutivity can be
proved also without using the $r$-matrix approach, just by
referring to Proposition \ref{invol}.

\subsubsection{$N=2$, the Lagrange top} \label{lagrange}

Fixing $N=2$ in the formulae of the previous Subsection we recover
the well-known dynamics of the three-dimensional Lagrange top
described in the rest frame \cite{A,BS,GZ,KPR,RSTS,Su}. In other
words the Lagrange top is the first element of the rational
one-body $\su2$ tower.

The Lagrange case of the rigid body motion around a fixed point in
a homogeneous field is characterized by the following data: the
inertia tensor is given by ${\rm diag} (1,1,I_3)$, $I_3 \in
\mathbb{R}$, which means that the body is rotationally symmetric
w.r.t. the third coordinate axis, and the fixed point lies on the
symmetry axis.

The equations of motion (in the rest frame) are given by: \beq
\dot {{\bf{z}}}_0 = [\,{\bf{p}},{\bf{z}}_{1}\,], \qquad \dot
{{\bf{z}}}_1 = [\,{\bf{z}}_0 , {\bf{z}}_1\, ], \label{1000y} \eeq
where ${\bf{z}}_0\in \mathbb{R}^3$ is the vector of kinetic
momentum of the body, ${\bf{z}}_1\in \mathbb{R}^3$ is the vector
pointing from the fixed point to the center of mass of the body
and ${\bf{p}}$ is the constant vector along the gravity field. An
external observer is mainly interested in the motion of the
symmetry axis of the top on the surface $\langle \, {\bf{z}}_1,
{\bf{z}}_1\, \rangle$=constant.

A remarkable feature of the equations of motion (\ref{1000y}) is
that they do not depend explicitly on the anisotropy parameter
$I_3$ of the inertia tensor \cite{BS}. Moreover they are
Hamiltonian equations w.r.t. the Lie-Poisson brackets on
$\mathfrak{e}(3)^*$, see Eq. (\ref{E3}).

The Hamiltonian function that generates the equations of motion
(\ref{1000y}) is given by \beq H_1^{(2)} = \langle \,{\bf{p}},
{\bf{z}}_{1} \, \rangle + \half \langle \, { \bf{z}}_0,
{\bf{z}}_{0} \, \rangle, \label{Hlag} \eeq and the complete
integrability of the model is ensured by the second integral of
motion $H_0^{(2)} = \langle \,{\bf{p}}, {\bf{z}}_{0} \, \rangle$.
These involutive Hamiltonians can be obtained using Eq. (\ref{34})
with $N=2$, namely considering the spectral invariants of the Lax
matrix  $\CL_2 (\l)$, see Eq. (\ref{LP7}). The remaining two
spectral invariants are given by the Casimir functions (\ref{67}).

\subsubsection{$N=3$, the first extension of the Lagrange top} \label{firstext}

Let us now consider the dynamical system governed by the Lax
matrix (\ref{LP7}) with $N=3$. The Lie-Poisson brackets are
explicitly given in Eqs. (\ref{e3p1},\ref{e3p2}). According to Eq.
(\ref{34}) the involutive Hamiltonians are:
$$
H_{0}^{(3)} = \langle \,{\bf{p}},  {\bf{z}}_0 \, \rangle , \qquad
 H_1^{(3)} = \langle \,{\bf{p}},  {\bf{z}}_1 \, \rangle+
 \half \langle \, {\bf{z}}_0,  {\bf{z}}_0 \, \rangle, \qquad
 H_2^{(3)} = \langle \,{\bf{p}},  {\bf{z}}_2 \, \rangle +
        \langle \,  {\bf{z}}_0,  {\bf{z}}_1 \, \rangle.
$$

Looking at the brackets (\ref{e3p1},\ref{e3p2}) and taking into
account that $ {\bf{z}}_0$ and $ {\bf{z}}_2$ span respectively
$\mathfrak{su}(2)^*$ and $\mathbb{R}^3$, we may interpret them as
the total angular momentum of the system and the vector pointing
from a fixed point (which we shall take as $(0,0,0) \in
\mathbb{R}^3$) to the centre of mass of a Lagrange top. Let us
remark that $ {\bf{z}}_0$ does not coincide with the angular
momentum of the top due to the presence of the vector
${\bf{z}}_1$. We think of ${\bf{z}}_1$, whose norm is not
constant, as the position of the moving centre of mass of the
system composed by the Lagrange top  and a satellite, whose
position is described by ${\bf{z}}_1- {\bf{z}}_2$. Here we are
assuming that both bodies have unit masses. Notice that the
integral $ H_1^{(3)}$ formally coincides with the physical
Hamiltonian of the Lagrange top (\ref{Hlag}) where  now the vector
${\bf{z}}_0$ is the angular momentum of system and the vector
${\bf{z}}_1$ describes the motion of the total centre of mass.

According to Eq. (\ref{1}) the Hamiltonian flow generated by the integral
$H_1^{(3)}$ reads
\beq
\dot { {\bf{z}} }_0 = [\,{\bf{p}},  {\bf{z}}_1\,], \qquad
\dot{  {\bf{z}} }_1 = [\,  {\bf{z}}_0,  {\bf{z}}_1\,] +
[\,{\bf{p}},   {\bf{z}}_2\,] , \qquad
\dot{ {\bf{z}}  }_2 = [\, {\bf{z}}_0,  {\bf{z}}_2\,]. \nonumber
\eeq
We  see that the vector $ {\bf{z}}_1$ does not rotate rigidly, though ${\bf{z}}_2$
does.

\subsection{The rational many-body $\su2$ tower}

The rational {\it {many-body $\su2$ tower}} may be constructed
simply regarding the Lax matrix (\ref{LP7}) as the local matrix of
a chain of many, say $M$, copies of the Lie-Poisson structure
$\mathcal{C}_N(\mathfrak{su}(2)^*)$. Indeed the $r$-matrix
formulation (\ref{pb4}) ensures that the Lax matrix
 \beq
\CL_{M,N}(\l) = {\bf{p}} +\sum_{k=1}^M \sum_{i=0}^{N-1} \frac{{\bf
z}_{i,k}}{(\l -\mu_k)^{i+1}}, \nonumber
 \eeq
with pairwise distinct poles $\mu_k$ of order $N$ describes an
integrable system defined on
$\oplus^M\mathcal{C}_N(\mathfrak{su}(2)^*)$ with the same
$r$-matrix formulation (\ref{pb4}). See \cite{MPR,P} for further
details.

Let us consider the special case $N=2$, namely the Lie-Poisson
algebra given by $\oplus^M\mathfrak{e}(3)^*$. The resulting
integrable system has been called {\it {Lagrange chain}} in
\cite{MPR,MPRS3}. We now present a new derivation of such a system
without using the $r$-matrix approach, but just considering the
contraction procedure of a rational $\su2$ Gaudin model defined on
$\oplus^{2M}\mathfrak{su}(2)^*$. According to Proposition \ref{ji}
the contraction of the direct sum of $2M$ copies of
$\mathfrak{su}(2)^*$ (i.e. $N=2$) leads to the Lie-Poisson
brackets on $\oplus^M\mathfrak{e}(3)^*$.

It is convenient to simplify the notation: \beq {\bf z}_{0,k} =
{\bf m}_k, \qquad {\bf z}_{1,k} = {\bf a}_k, \qquad 1\leq k \leq
M. \label{uu} \eeq We interpret ${\bf{m}}_k =( m_k^1,  m_k^2,
m_k^3) \in \mathbb{R}^3$ and ${\bf{a}}_k =( a_k^1,  a_k^2, a_k^3)
\in \mathbb{R}^3$ as, respectively, the angular momentum and the
vector pointing from the fixed point to the center of mass of the
$k$-th top. The Lie-Poisson bracket on $\oplus^M
\mathfrak{e}(3)^*$ is: \beq \left\{ m_k^{\alpha}, m_j ^{\beta}
\right\}=- \delta_{k,j} \, \ep_{\al \be \gamma} \,m_k^{\gamma},
\qquad \left\{ m_k^{\alpha}, a_j ^{\beta}   \right\}=-
\delta_{k,j} \, \ep_{\al \be \gamma} \,a_k^{\gamma}, \qquad
\left\{ a_k^{\alpha}, a_j ^{\beta} \right\}= 0, \label{xc} \eeq
with $1 \leq k,j \leq M$. This bracket possesses $2M$ Casimir
functions: \beq Q_k^{(1)} = \langle \, {\bf{m}}_k, {\bf{a}}_k \,
\rangle,  \qquad Q_k^{(2)} = \half \langle \, {\bf{a}}_k,
{\bf{a}}_k \, \rangle, \qquad 1 \leq k \leq M. \label{ds3} \eeq
Using the notation introduced in Eq. (\ref{uu}), the Lax matrix of
the Lagrange chain reads \beq \CL_{M,2}(\l) =  {\bf{p}} +
\sum_{i=1}^{M} \left[\, \frac{{\bf{m}}_i}{\l - \mu_i} +
\frac{{\bf{a}}_i}{(\l - \mu_i)^2} \, \right].\label{lgausum4} \eeq

Let us now consider a rational $\su2$ Gaudin model with $2M$
poles. We have to apply  the map defined in Eq. (\ref{ty234}) to
the set of $\mathbb{R}^3$ vectors $\{{\bf y}_i \}_{i=1}^{2M}$:
\beq ({\bf z}_{0})_i = {\bf m}_i =  {\bf y}_{2i} +{\bf y}_{2i-1},
\qquad ( {\bf z}_{1})_i = {\bf a}_i = \vartheta \, (\nu_{2i} \,
{\bf y}_{2i} +\nu_{2i-1} \, {\bf y}_{2i-1}), \label{89} \eeq with
$1\leq i \leq M$. Moreover we define the following pole
coalescence: \beq \l_{2i} =  \mu_i + \vartheta \,\nu_{2i}, \qquad
\l_{2i-1} = \mu_i+ \vartheta \,\nu_{2i-1} , \qquad 1\leq i \leq
M,\label{mku} \eeq where the $\l_i$ are the $2M$ parameters of the
rational $\su2$ Gaudin model.

\begin{prop} \label{mb}

Consider the Lax equation (\ref{kh}) with the pole coalescence
(\ref{mku}). Under the map (\ref{ty2}) and upon the limit
$\vartheta \rightarrow 0$ it tends to \beq \dot \CL_{M,2} (\l)
=\left[\,  \CL_{M,2}(\l), \CM_{M,2}^{(-)} (\l)  \, \right]=
-\left[\,  \CL_{M,2}(\l), \CM_{M,2}^{(+)} (\l)  \,
\right],\label{saw} \eeq with the matrix $\CL_{M,2}(\l)$ given by
Eq. (\ref{lgausum4}) and \beq \CM_{M,2}^{(-)} (\l) =
\sum_{i=1}^{M} \frac{1}{\l-\mu_{i}} \left[\, \mu_i \, {\bf{m}}_i +
\frac{\l \, {\bf{a}}_i }{\l-\mu_{i}} \, \right], \qquad
\CM_{M,2}^{(+)} (\l) = \l \,{\bf{p}}  + \sum_{i=1}^{M}  {\bf{m}}_i
.  \label{mnb} \eeq

\end{prop}

{\bf{Proof:}} We have:
\bea
\CL_{\CG}(\l)  &=&
 {\bf{p}} + \sum_{i=1}^{M} \left(\frac{{\bf{y}}_{2i-1} }{\l - \mu_i- \vartheta\, \nu_{2i-1}}+
 \frac{{\bf{y}}_{2i} }{\l - \mu_i - \vartheta\, \nu_{2i}} \right)=
\nonumber \\
 &=& {\bf{p}} + \sum_{i=1}^{M} \frac{{\bf{y}}_{2i-1} + {\bf{y}}_{2i} }{\l - \mu_i}
+\sum_{i=1}^{M} \frac{\vartheta \, (\nu_{2i} \,  {\bf y}_{2i}
+\nu_{2i-1} \, {\bf y}_{2i-1})} {(\l - \mu_i)^2}+ O(\vartheta)
\xrightarrow{\scriptstyle {\vartheta \rightarrow 0}} \CL_{M,2}(\l)
. \nonumber \eea A similar computation leads to the auxiliary
matrices $\CM_{M,2}^{(\pm)}$ in Eq. (\ref{mnb}) starting from the
ones in Eq. (\ref{100g}).

\endpf

The Hamiltonian flow described by the Lax equation (\ref{saw}) is given by
\beq
\dot {\bf{m}}_i = \left[\,  {\bf{p}} , {\bf{a}}_i \, \right] +
\left[\, \mu_i \, {\bf{p}} +
{\textstyle \sum_{k=1}^M }  {\bf{m}}_k , {\bf{m}}_i \, \right] ,\qquad
\dot {\bf{a}}_i = \left[\,\mu_i \, {\bf{p}} + {\textstyle \sum_{k=1}^M }  {\bf{m}}_k
 , {\bf{a}}_i \, \right] \, , \label{666u}
\eeq with $1 \leq i \leq M$, while the characteristic equation
$\det( \CL_{M,2}(\l) -\mu \, {\bf 1})=0$ reads \beq
 -\mu^2= \frac{1}{4} \langle \,{\bf{p}}, {\bf{p}}
\, \rangle + \half \sum_{k=1}^M \left[\, \frac{R_k}{\l
-\mu_k}+\frac{S_k}{(\l -\mu_k)^2}+ \frac{Q_k^{(1)}}{(\l
-\mu_k)^3}+ \frac{Q_k^{(2)}}{(\l -\mu_k)^4} \, \right], \nonumber
\eeq where the functions $Q_{k}^{(1)}$, $Q_{k}^{(2)}$ are the
Casimir functions (\ref{ds3}), and the functions
\begin{eqnarray}
R_k & = & \langle \,{\bf{p}}, {\bf{m}}_k \, \rangle +
\sum_{\stackrel{\scriptstyle{j=1}} {j \neq k}}^M \left[\, \frac{
\langle \,{\bf{m}}_k, {\bf{m}}_j \, \rangle}{\mu_k - \mu_j}+
\frac{ \langle \,{\bf{m}}_k, {\bf{a}}_j \, \rangle  - \langle
\,{\bf{m}}_j, {\bf{a}}_k \, \rangle } {(\mu_k - \mu_j)^2}
 - 2 \, \frac{ \langle \,{\bf{a}}_k, {\bf{a}}_j \, \rangle }{(\mu_k - \mu_j)^3}
\, \right], \quad \label{567} \\
S_k & = & \langle \,{\bf{p}}, {\bf{a}}_k\, \rangle + \half \langle
\,{\bf{m}}_k, {\bf{m}}_k \, \rangle +
\sum_{\stackrel{\scriptstyle{j=1}} {j \neq k}}^M \left[\, \frac{
\langle \, {\bf{a}}_k, {\bf{m}}_j \, \rangle }{\mu_k - \mu_j}+
 \frac{ \langle \,{\bf{a}}_k, {\bf{a}}_j \, \rangle }{(\mu_k - \mu_j)^2}
\, \right], \label{568}
\end{eqnarray}
are the $2M$ independent and involutive Hamiltonians of the Lagrange chain.

Notice that, as in the $\su2$ rational Gaudin model, there is a
linear integral given by $ \sum_{k=1}^M  R_k= \sum_{k=1}^M \langle
\,{\bf{p}},{\bf{m}}_k \, \rangle. \nonumber $ A possible choice
for a physical Hamiltonian describing the dynamics of the model
can be constructed considering a linear combination of the
Hamiltonians $R_{k}$ and $S_{k}$ similar to the one considered in
Eq. (\ref{itglrrrr}). We have: \beq \CH_{M,2} = \sum_{k=1}^M (
\mu_k \, R_k + S_k)=
 \sum_{k=1}^M \langle \,{\bf{p}}, \mu_k \, {\bf{m}}_k + {\bf{a}}_k\, \rangle+
\half \sum_{i,k=1}^M \langle \,{\bf{m}}_i, {\bf{m}}_k \, \rangle.
\label{H!} \eeq
It is easy to check that the integral (\ref{H!})
generates the Hamiltonian flow (\ref{666u}). If $M=1$, the
Hamiltonian (\ref{H!}) gives the sum of the two integrals of
motion of the Lagrange top.

We can construct the integrals of motion of the Lagrange chain
also by using the Lie-Poisson map (\ref{ty234}) with the pole
coalescence (\ref{mku}) directly in the Hamiltonians (\ref{35}),
according to
 \beq
R_{i} = \lim_{\vartheta \rightarrow 0} \, [\,H_{2i} + H_{2i-1}\,],
\qquad  S_{i}= \lim_{\vartheta \rightarrow 0}\, [\, \vartheta \,
(\nu_{2i} \,  H_{2i} +\nu_{2i-1} \, H_{2i-1})\,].\eeq

\section{Discrete-time rational $\su2$ Gaudin models} \label{dgm} \label{sec4}

The main goal of this Section is the construction of an integrable
Poisson map discretizing the Hamiltonian flow (\ref{1g}). We shall
provide an explicit map approximating, for a small discrete-time
step $\ep$, the time $\ep$ shift along the trajectories of the
equations of motion (\ref{1g}) generated by the Hamiltonian
function (\ref{itglt}). We have to remark that no Lax
representation (hence no $r$-matrix formulation) has been found
for this map. Its Poisson property and integrability will be
proved by direct inspection.

\begin{prop} \label{popo}
The map
\beq
\CD_\ep^{N}: \; {{\bf{y}}}_i  \mapsto \widehat {{\bf{y}}}_i =
\left( {\bf 1}+ \ep \, \l_i \, {{\bf{p}}} \right)
\left({\bf 1}+ \ep \, {\textstyle \sum_{j=1}^N }  {{\bf{y}}}_j \right)
\,{{\bf{y}}}_i \,
\left({\bf 1}+ \ep \, {\textstyle \sum_{j=1}^N }  {{\bf{y}}}_j \right)^{-1}
\left( {\bf 1}+ \ep \, \l_i \, {{\bf{p}}} \right)^{-1}, \label{mg43}
\eeq
with $1 \leq i \leq N$ and $\ep \in \mathbb{R}$,
is Poisson w.r.t. the brackets (\ref{LPgdd}) on $\oplus^N \mathfrak{su}(2)^*$
and has $N$ independent and involutive integrals of motion assuring
its complete integrability:
\beq
H_k(\ep) = \langle \,  {\bf{p}} ,  {\bf{y}}_k \, \rangle +
\sum_{\stackrel{\scriptstyle{j=1}} {j \neq k}}^N
\frac{\langle \, {\bf{y}}_k,{\bf{y}}_j \, \rangle }{\l_k - \l_j}
\left( 1 +\frac{\ep^2}{4}  \l_k \, \l_j \, \langle \, {\bf{p}}  ,{\bf{p}} \, \rangle   \right)
-\frac{\ep}{2} \sum_{\stackrel{\scriptstyle{j=1}} {j \neq k}}^N
\, \langle \, {\bf{p}}  , [\,{\bf{y}}_k ,{\bf{y}}_j ]\, \rangle, \label{disc}
\eeq
with $1 \leq k \leq N$.
\end{prop}

{\bf{Proof:}} Let us first  notice that the map (\ref{mg43})
reproduces at order $\ep$ the continuous-time Hamiltonian flow
(\ref{1g}). The map (\ref{mg43}) is the composition of two
non-commuting conjugations:
$
\CD_\ep^{N} = (\CD_\ep^{N})_2 \circ (\CD_\ep^{N})_1,
$
where
\begin{eqnarray}
(\CD_\ep^{N})_1: &\;& {\bf{y}}_i \mapsto  {\bf{y}}_i^* =
\left({\bf 1}+\ep \, \textstyle{\sum_{j=1}^N} {\bf{y}}_j \right)
\,{\bf{y}}_i \,
\left({\bf 1}+\ep \,\textstyle{\sum_{j=1}^N} {\bf{y}}_j\right)^{-1}, \label{q1} \\
\nonumber\\ (\CD_\ep^{N})_2: &\;& {\bf{y}}_i^* \mapsto \widehat
{\bf{y}}_i = \left( {\bf 1}+ \ep \, \l_i \, {\bf{p}} \right)
{\bf{y}}^*_i \left({\bf 1}+ \ep \, \l_i \, {\bf{p}} \right)^{-1},
\label{q2}
\end{eqnarray}
with $1 \leq i \leq N$. Notice that $(\CD_\ep^{N})_1 \circ (\CD_\ep^{N})_2
\neq (\CD_\ep^{N})_2 \circ (\CD_\ep^{N})_1$.

The Poisson property of the map $\CD_\ep^{N}$ is a consequence of
the Poisson property of the maps $(\CD_\ep^{N})_1$ and
$(\CD_\ep^{N})_2$. In fact $(\CD_\ep^{N})_1$ is a shift along a
Hamiltonian flow on $\oplus^N \mathfrak{su}(2)^*$ w.r.t. the
Hamiltonian $\sum_{j\neq k=1}^N\langle \, {\bf{y}}_j,{\bf{y}}_k \,
\rangle$. On the other hand  $(\CD_\ep^{N})_2$ is a shift along a
Hamiltonian flow on $\oplus^N \mathfrak{su}(2)^*$ w.r.t. the
Hamiltonian $\sum_{k=1}^N \langle \, {\bf{p}},\l_k \,
{\bf{y}}_k^*\, \rangle.$ Therefore the composition
$(\CD_\ep^{N})_2 \circ (\CD_\ep^{N})_1$ is a Poisson map w.r.t.
the bracket (\ref{LPgdd}).

Let us now prove the complete integrability of the map
(\ref{mg43}). We show that the functions (\ref{disc}) are indeed
integrals of the map (\ref{mg43}). Their independence is clear,
while their involution w.r.t. the brackets (\ref{LPgdd}) is proved
in Appendix 2.

Notice that the maps (\ref{q1}), (\ref{q2}) imply, respectively,
the following relations:
\begin{eqnarray}
  \langle \, {\bf{y}}_i^*, {\bf{y}}_j^* \, \rangle =
\langle \, {\bf{y}}_i, {\bf{y}}_j \, \rangle, & \quad &
{\bf{y}}_i^* + \frac{\ep}{2} \sum_{j=1}^N \left[\,  {\bf{y}}_i^* ,
{\bf{y}}_j \, \right] = {\bf{y}}_i + \frac{\ep}{2} \sum_{j=1}^N
\left[\,{\bf{y }}_j , {\bf{y}}_i \, \right], \label{qq1} \\
\nonumber\\
 \langle \, {\bf{p}}, \hat {\bf{y}}_j \, \rangle = \langle \,
{\bf{p}}_i, {\bf{y}}_j^* \, \rangle, & \quad &  \hat {\bf{y}}_i +
\frac{\ep}{2} \l_i \, \left[ \,\hat {\bf{y}}_i , {\bf{p}}\,
\right] = {\bf{y}}_i^* + \frac{\ep}{2} \, \l_i  \, \left[\,
{\bf{p}} , {\bf{y}}_i^*\, \right] , \label{qq2}
\end{eqnarray}
with $1\leq i,j \leq N$. The preservation of the functions (\ref{disc}) is
demonstrated by the following computation:
\bea
\hat H_k(\ep) &=&  \langle \,  {\bf{p}} ,  \hat {\bf{y}}_k \, \rangle + \sum_{\stackrel{\scriptstyle{j=1}} {j \neq k}}^N
\frac{\langle \, \hat {\bf{y}}_k,\hat {\bf{y}}_j \, \rangle }{\l_k - \l_j}
\left( 1 +\frac{\ep^2}{4}  \l_k \, \l_j \, \langle \, {\bf{p}}  ,{\bf{p}} \, \rangle   \right)
-\frac{\ep}{2} \sum_{\stackrel{\scriptstyle{j=1}} {j \neq k}}^N
\, \langle \, {\bf{p}}  , [\, \hat {\bf{y}}_k ,\hat {\bf{y}}_j ]\, \rangle = \nonumber \\
&=&\langle \,  {\bf{p}} ,  {\bf{y}}_k^* \, \rangle + \sum_{\stackrel{\scriptstyle{j=1}} {j \neq k}}^N
\frac{\langle \, {\bf{y}}_k^* , {\bf{y}}_j^*  \, \rangle }{\l_k - \l_j}
\left( 1 +\frac{\ep^2}{4}  \l_k \, \l_j \, \langle \, {\bf{p}}  ,{\bf{p}} \, \rangle   \right)
+\frac{\ep}{2} \sum_{\stackrel{\scriptstyle{j=1}} {j \neq k}}^N
\, \langle \, {\bf{p}}  , [ \, {\bf{y}}_k^*  , {\bf{y}}_j^*  ]\, \rangle= \nonumber \\
&=&\langle \,  {\bf{p}} ,  {\bf{y}}_k\, \rangle + \sum_{\stackrel{\scriptstyle{j=1}} {j \neq k}}^N
\frac{\langle \, {\bf{y}}_k, {\bf{y}}_j  \, \rangle }{\l_k - \l_j}
\left( 1 +\frac{\ep^2}{4}  \l_k \, \l_j \, \langle \, {\bf{p}}  ,{\bf{p}} \, \rangle   \right)
-\frac{\ep}{2} \sum_{\stackrel{\scriptstyle{j=1}} {j \neq k}}^N
\, \langle \, {\bf{p}}  , [\,  {\bf{y}}_k  , {\bf{y}}_j ]\, \rangle = H_k(\ep), \nonumber
\eea
with $1 \leq k \leq N$. Here we have used Eq. (\ref{qq2}) in the first step and Eq. (\ref{qq1}) in the second one.

\endpf

Using the discrete Hamiltonians (\ref{disc}) we can compute the
discrete-time version of the Hamiltonian (\ref{itglt}). It reads:
\bea
\CH_{\CG}(\ep) = \sum_{k=1}^N \l_k \,H_k(\ep) &=&
\sum_{k=1}^N \langle \,  {\bf{p}} , \l_k \, {\bf{y}}_k \, \rangle+
\half \sum_{\stackrel{\scriptstyle{j,k=1}} {j \neq k}}^N \langle \,  {\bf{y}}_k , {\bf{y}}_j \, \rangle
\left( 1 +\frac{\ep^2}{4} \, \l_k \, \l_j  \,\langle \, {\bf{p}}  ,{\bf{p}} \, \rangle\right) - \nonumber \\
&& -\, \frac{\ep}{4} \sum_{\stackrel{\scriptstyle{j,k=1}} {j \neq
k}}^N (\l_k -\l_j) \, \, \langle \, {\bf{p}}  ,  [\,{\bf{y}}_k ,
{\bf{y}}_j \,]\, \rangle. \nonumber \eea Moreover we still have a
linear integral given by $\sum_{k=1}^N  H_k(\ep) = \sum_{k=1}^N
\langle \,  {\bf{p}} , {\bf{y}}_k \, \rangle$, as in the
continuous-time case.

\section{Contractions  of discrete-time rational $\su2$ Gaudin models} \label{sec5}

Performing the contraction procedures presented in Subsections
\ref{p5} and \ref{p6} we can now construct the integrable
discrete-time versions of the Hamiltonian flows (\ref{1}) and
(\ref{666u}) of the whole rational one-body $\su2$ tower and of
the Lagrange chain.

\subsection{The discrete-time one-body $\su2$ tower}

The integrable Poisson map discretizing the flow (\ref{1}) of the
rational one-body $\su2$ tower is given in the following
Proposition.

\begin{prop} \label{popore}
The map
\beq
 \til {\CD}_\ep^{N}: \;  {\bf{z}}_i  \mapsto {\hat { {\bf{z}}}}_i =
({\bf 1}+\ep \,  {\bf{z}}_0) \, {\bf{z}}_i \,
({\bf 1}+\ep \, {\bf{z}}_0)^{-1}- 2
\sum_{j=1}^{N-i-1} \left( -\frac{\ep}{2}\right)^j {\rm ad}_{\bf{p}}^j \, {\hat { {\bf{z}}}}_{i+j},
 \label{mgtilde}
\eeq
with $0 \leq i \leq N-1$ and $\ep \in \mathbb{R}$,
is Poisson w.r.t. the brackets (\ref{np2}) on $\mathcal{C}_N(\mathfrak{su}(2)^*)$
and has $N$ independent  and involutive integrals of motion assuring
its complete integrability:
\beq
{H}_k^{(N)} (\ep) = \langle \,{\bf{p}}, { {\bf{z}}}_{k} \, \rangle +
\half \sum_{i=0}^{k-1} \langle \,  {{ \bf{z}}}_i, { {\bf{z}}}_{k-i-1} \, \rangle
+ \frac{\ep}{2} \langle \, {\bf{p}}  ,   [\,{{\bf{z}}}_0 ,  {{\bf{z}}}_k ]\, \rangle
+\frac{\ep^2}{8}\langle \,{\bf{p}}, {\bf{p}}  \, \rangle
\sum_{i=0}^{k-1} \langle \, { { \bf{z}}}_{i+1}, {  {\bf{z}}}_{k-i} \, \rangle, \label{ggh}
\eeq
with $0 \leq k \leq N-1$.
\end{prop}

{\bf{Proof:}} Let us construct the map (\ref{mgtilde}) through the
usual contraction procedure and the pole coalescence $\l_i
=\vartheta \, \nu_i$, $1 \leq i \leq N$, performed on the map
(\ref{mg43}).

Consider the map $(\CD_\ep^{N})_1$ in Eq. (\ref{q1}). Using the map (\ref{ty2})
and assuming $\l_i =\vartheta \, \nu_i$, $1 \leq i \leq N$, we get:
\bea
 {\bf{z}}_i^* =
\sum_{k=1}^N \vartheta^i \, \nu_k^i \, {\bf{y}}_k^* &=&
\sum_{k=1}^N \vartheta^i \, \nu_k^i \, \left({\bf 1}+\ep \,
\textstyle{\sum_{j=1}^N} {\bf{y}}_j \right) \,{\bf{y}}_k \,
\left({\bf 1}+\ep \,\textstyle{\sum_{j=1}^N} {\bf{y}}_j\right)^{-1} =
\nonumber \\
&=& ({\bf 1}+\ep \,  {\bf{z}}_0) \, {\bf{z}}_i \,
({\bf 1}+\ep \, {\bf{z}}_0)^{-1}, \nonumber
\eea
with $0 \leq i \leq N-1$.
Hence the contracted version of $(\CD_\ep^{N})_1$ is given by
\beq
( \til \CD_\ep^{N})_1: \;  {\bf{z}}_i \mapsto   {\bf{z}}_i^*
= ({\bf 1}+\ep \,  {\bf{z}}_0) \, {\bf{z}}_i \,
({\bf 1}+\ep \, {\bf{z}}_0)^{-1}, \qquad 0 \leq i \leq N-1. \nonumber
\eeq

On the other hand, a direct computation, with the help of Eq.
(\ref{lemm1}), yields the contracted version of the map
$(\CD_\ep^{N})_2$ in Eq. (\ref{q2}): \bea \sum_{k=1}^N \vartheta^i
\, \nu_k^i \, \hat {\bf{y}}_k &=& \sum_{k=1}^N \vartheta^i \,
\nu_k^i \, \left( {\bf 1}+ \ep \, \vartheta \, \nu_k \, {\bf{p}}
\right) {\bf{y}}^*_k
\left({\bf 1}+ \ep \, \vartheta \, \nu_k \, {\bf{p}} \right)^{-1}  = \nonumber \\
&=& \sum_{k=1}^N \sum_{j \geq 0} \vartheta^{i+j} \, \nu_k^{i+j} \, (-\ep)^j \,
\left( {\bf 1}+ \ep \, \vartheta \, \nu_k \, {\bf{p}} \right) \, {\bf{y}}^*_k \,
{\bf{p}}^j= \nonumber \\
&=& {{ {\bf{z}}}}_{i}^* +
2 \sum_{j=1}^{N-i-1} \left( \frac{\ep}{2}\right)^j {\rm ad}_{\bf{p}}^j \,
{{ {\bf{z}}}}_{i+j}^* + O(\vartheta), \nonumber
\eea
with $0 \leq i \leq N-1$.
Performing the limit $\vartheta \rightarrow 0$ we have:
\beq
( \til \CD_\ep^{N})_2: \;  {\bf{z}}^*_i \mapsto  \hat {{\bf{z}}}_i={{ {\bf{z}}}}_{i}^* +
2 \sum_{j=1}^{N-i-1} \left( \frac{\ep}{2}\right)^j {\rm ad}_{\bf{p}}^j \, {{ {\bf{z}}}}_{i+j}^*,
\qquad 0 \leq i \leq N-1.
\nonumber
\eeq

Now the composition $( \til \CD_\ep^{N})_2 \circ ( \til
\CD_\ep^{N})_1$ is easily verified to result in the map $\til
\CD_\ep^{N}$ given in Eq. (\ref{mgtilde}). The Poisson property of
the map $ \til \CD_\ep^{N}$ is a consequence of the one of the map
$\CD_\ep^{N}$ in Eq. (\ref{mg43}).

Next, we construct, by contraction of the functions (\ref{disc}),
the integrals of the Poisson map (\ref{mgtilde}). We know that
fixing $\ep=0$ in Eq. (\ref{disc}) we recover the Hamiltonians
(\ref{35}) of the continuous-time $\mathfrak{su}(2)$ rational
Gaudin model. Their contraction gives the Hamiltonians (\ref{34})
of the continuous-time rational one-body $\su2$ tower. Therefore
it is enough to perform the contraction procedure just on the two
$\ep$-dependent terms of the integrals (\ref{disc}). We have: \bea
\sum_{k=1}^{N} \vartheta^i \, \nu_k^i \, H_k (\ep)&=& \langle
\,{\bf{p}},  {\bf{z}}_{i} \, \rangle + \half \sum_{m=0}^{i-1}
\langle \,  { \bf{z}}_m,  {\bf{z}}_{i-m-1} \, \rangle - \nonumber \\
&& - \frac{\ep}{4} \sum_{\stackrel{\scriptstyle{j,k=1}} {j \neq k}}^N (\vartheta^i \, \nu_k^i - \vartheta^i \, \nu_j^i )
\langle \, {\bf{p}}, [\,{\bf{y}}_{k} ,{\bf{y}}_{j}] \, \rangle + \nonumber \\
&& +\frac{\ep^2}{8}\langle \,{\bf{p}}, {\bf{p}}  \, \rangle  \sum_{\stackrel{\scriptstyle{j,k=1}} {j \neq k}}^N \vartheta^{i+1}
\frac{\nu_k^{i+1} \, \nu_j - \nu_j^{i+1} \, \nu_k }{\nu_k - \nu_j} \langle \, {\bf{y}}_k,{\bf{y}}_j \, \rangle = \nonumber \\
&=& \langle \,{\bf{p}},  {\bf{z}}_{i} \, \rangle + \half  \sum_{m=0}^{i-1}
\langle \,  { \bf{z}}_m,  {\bf{z}}_{i-m-1} \, \rangle + \frac{\ep}{2}
\langle \, {\bf{p}}, [\, {\bf{z}}_{0} ,  {\bf{z}}_{i} \, ]\, \rangle + \nonumber \\
&& +\frac{\ep^2}{8}\langle \,{\bf{p}}, {\bf{p}}  \, \rangle  \sum_{m=0}^{i-1} \sum_{\stackrel{\scriptstyle{j,k=1}} {j \neq k}}^N
(\vartheta \, \nu_k)^{m+1} (\vartheta \, \nu_j)^{i-m} \langle \, {\bf{y}}_k,{\bf{y}}_j \, \rangle = {H}_i^{(N)} (\ep), \nonumber
\eea
with $0 \leq i \leq N-1$.

The involutivity of the integrals $\{{H}_k^{(N)}(\ep) \}_{k=0}^{N-1}$ is ensured
thanks to Proposition \ref{invol}.
\endpf

Let us remark that the specialization to $N=2$ of the map
(\ref{mgtilde}) gives the integrable time-discretization of the
Lagrange top found by A.I. Bobenko and Yu.B. Suris in \cite{BS}.
According to Eq. (\ref{mgtilde}) it reads: \beq \hat {{\bf{z}}}_0
= {{\bf{z}}}_0  + \ep \, \left[\,{\bf{p}} , \hat {{\bf{z}}}_{1}\,
\right], \qquad \hat {{\bf{z}}}_1 = ({\bf 1} + \ep \,
{{\bf{z}}}_0) \, {{\bf{z}}}_1 \, ({\bf 1} + \ep \,
{{\bf{z}}}_0)^{-1}. \label{1000ydr} \eeq The above explicit map
approximates, for small $\ep$, the time $\ep$ shift along the
trajectories of the Hamiltonian flow (\ref{1000y}). This
distinguish the situation from the map in  \cite{MV}, where
Lagrangian equations led to correspondences rather than to maps.

The map (\ref{1000ydr}) is Poisson w.r.t. the bracket (\ref{E3})
on $\mathfrak{e}(3)^*$ and its complete integrability is ensured
by the integrals of motion \beq H_0^{(2)} = \langle \,{\bf{p}},
{\bf{z}}_{0} \, \rangle, \qquad H_1^{(2)}(\ep) = \half \langle \,
{\bf{z}}_0,  {\bf{z}}_{0} \, \rangle + \langle \,{\bf{p}},
{\bf{z}}_{1} \, \rangle + \frac{\ep}{2} \langle \,{\bf{p}},
\left[\,{\bf{z}}_{0}, {\bf{z}}_{1}\, \right] \, \rangle. \eeq

A remarkable feature of the map (\ref{1000ydr}) is that it admits
a Lax representation and the same linear $r$-matrix bracket
(\ref{pb4}) as in the continuous case, see \cite{BS} for further
details. The Lax matrix of the map is a deformation of the Lax
matrix of the Lagrange top.

\subsection{The discrete-time Lagrange chain}

The integrable Poisson map discretizing the flow (\ref{666u}) of
the Lagrange chain is given in the following Proposition.

\begin{prop}
The map
\begin{subequations}
\begin{align}
& \! \! \! \!\!\!\!\! \widehat {{\bf{m}}}_i =
\left( {\bf 1}+ \ep \, \mu_i \, {{\bf{p}}} \right)
\left({\bf 1}+ \ep \, {\textstyle \sum_{j=1}^M }  {{\bf{m}}}_j \right)
\,{{\bf{m}}}_i \,
\left({\bf 1}+ \ep \, {\textstyle \sum_{j=1}^M }  {{\bf{m}}}_j \right)^{-1}
\left( {\bf 1}+ \ep \, \mu_i \, {{\bf{p}}} \right)^{-1} + \ep \, \left[\,  {{\bf{p}}}, \widehat {{\bf{a}}}_i\, \, \right], \label{tra} \\
& \! \! \! \!\!\!\!  \widehat {{\bf{a}}}_i =
\left( {\bf 1}+ \ep \, \mu_i \, {{\bf{p}}} \right)
\left({\bf 1}+ \ep \, {\textstyle \sum_{j=1}^M }  {{\bf{m}}}_j \right)
\,{{\bf{a}}}_i \,
\left({\bf 1}+ \ep \, {\textstyle \sum_{j=1}^M }  {{\bf{m}}}_j \right)^{-1}
\left( {\bf 1}+ \ep \, \mu_i \, {{\bf{p}}} \right)^{-1} , \label{trb}
\end{align}
\end{subequations}
with $1 \leq i \leq M$ and $\ep \in \mathbb{R}$,
is Poisson w.r.t. the brackets (\ref{xc})  on $\oplus^M \mathfrak{e}(3)^*$
and has $2M$ independent and involutive integrals of motion assuring
its complete integrability:
\begin{eqnarray}
  R_k(\ep)  & = &  \langle \,{\bf{p}}, {\bf{m}}_k \, \rangle
-\frac{\ep}{2} \langle \,{\bf{p}}, \left[\,{\bf{m}}_k ,
{\textstyle \sum_{j=1}^M }  {{\bf{m}}}_j \, \, \right] \, \rangle \nonumber \\
 & & \hspace{-1.5cm} +  \sum_{\stackrel{\scriptstyle{j=1}} {j \neq k}}^M
\left[\, \left( \frac{ \langle \,{\bf{m}}_k, {\bf{m}}_j \,
\rangle}{\mu_k - \mu_j} - 2 \, \frac{ \langle \,{\bf{a}}_k,
{\bf{a}}_j \, \rangle }{(\mu_k - \mu_j)^3}\right) \, \left(1 +
\frac{\ep^2}{4} \, \mu_k \, \mu_j \,\langle \,{\bf{p}}, {\bf{p}}
\, \rangle   \right) \right.
\nonumber \\
 & & \hspace{-1.5cm} + \left. \frac{ \langle \,{\bf{m}}_k, {\bf{a}}_j \, \rangle }
{(\mu_k - \mu_j)^2} \, \left(1 + \frac{\ep^2}{4} \, \mu_k^2
\,\langle \,{\bf{p}}, {\bf{p}} \, \rangle \right) - \frac{ \langle
\,{\bf{m}}_j, {\bf{a}}_k \, \rangle } {(\mu_k - \mu_j)^2} \,
\left(1 + \frac{\ep^2}{4} \, \mu_j^2  \,\langle \,{\bf{p}},
{\bf{p}} \, \rangle \right)
\, \right], \label{567dd} \\
 \nonumber \\
 S_k(\ep) & = &  \langle \,{\bf{p}}, {\bf{a}}_k\, \rangle +
 \half \langle \,{\bf{m}}_k, {\bf{m}}_k \, \rangle \, \left(1+
\frac{\ep^2}{4} \, \mu_k^2  \,\langle \,{\bf{p}}, {\bf{p}} \, \rangle \right)
-\frac{\ep}{2} \langle \,{\bf{p}}, \left[\,{\bf{a}}_k ,
{\textstyle \sum_{j=1}^M }  {{\bf{m}}}_j \, \, \right] \, \rangle
 \nonumber \\
& & \hspace{-1.5cm} +\sum_{\stackrel{\scriptstyle{j=1}} {j \neq
k}}^M \left[\, \frac{ \langle \, {\bf{a}}_k, {\bf{m}}_j \, \rangle
}{\mu_k - \mu_j}\, \left(1+ \frac{\ep^2}{4} \, \mu_k \, \mu_j
\,\langle \,{\bf{p}}, {\bf{p}} \, \rangle \right) +
 \frac{ \langle \,{\bf{a}}_k, {\bf{a}}_j \, \rangle }{(\mu_k - \mu_j)^2} \, \left(1+
\frac{\ep^2}{4} \, \mu_k^2  \,\langle \,{\bf{p}}, {\bf{p}} \,
\rangle \right) \, \right], \quad\label{568dd}
\end{eqnarray}
with $1 \leq k \leq M$.
\end{prop}

{\bf{Proof:}}
Using the map (\ref{89}) and the pole coalescence (\ref{mku})
in the map (\ref{q1}) with $N = 2M$ we immediately obtain
the contracted version of $(\CD_\ep^{2M})_1$. It reads
\begin{subequations}
\begin{align}
& {\bf m}_i^* = {\bf y}_{2i}^* +{\bf y}_{2i-1}^*=
\left({\bf 1}+ \ep \, {\textstyle \sum_{j=1}^M }  {{\bf{m}}}_j \right)
\,{{\bf{m}}}_i \,
\left({\bf 1}+ \ep \, {\textstyle \sum_{j=1}^M }  {{\bf{m}}}_j \right)^{-1}, \label{tr1}\\
& {\bf a}_i^* = \vartheta \, (\nu_{2i} \,  {\bf y}_{2i}^* +\nu_{2i-1} \, {\bf y}_{2i-1}^*)=
\left({\bf 1}+ \ep \, {\textstyle \sum_{j=1}^M }  {{\bf{m}}}_j \right)
\,{{\bf{a}}}_i \,
\left({\bf 1}+ \ep \, {\textstyle \sum_{j=1}^M }  {{\bf{m}}}_j \right)^{-1},\label{tr2}
\end{align}
\end{subequations}
with $1 \leq i \leq M$. The same procedure leads to
the contracted version of $(\CD_\ep^{2M})_2$. It reads
\begin{subequations}
\begin{align}
\hat {\bf m}_i &= \hat {\bf y}_{2i} +\hat {\bf y}_{2i-1}= \left({\bf 1}+ \ep \, \mu_i \, {{\bf{p}}} \right)
\,{{\bf{m}}}_i^* \,
\left({\bf 1}+ \ep \, \mu_i \,{{\bf{p}}} \right)^{-1} +\nonumber \\
& \quad + \ep \, \left[\, {{\bf{p}}},
\left({\bf 1}+ \ep \, \mu_i \, {{\bf{p}}} \right)
\,{{\bf{a}}}_i^* \,
\left({\bf 1}+ \ep \, \mu_i \,{{\bf{p}}} \right)^{-1} \, \right]
+ O(\vartheta), \label{tr3} \\
 \hat {\bf a}_i &= \vartheta \, (\nu_{2i} \,
 \hat {\bf y}_{2i} +\nu_{2i-1} \, \hat {\bf y}_{2i-1})=
\left({\bf 1}+ \ep \, \mu_i \, {{\bf{p}}} \right)
\,{{\bf{a}}}_i^* \,
\left({\bf 1}+ \ep \, \mu_i \,{{\bf{p}}} \right)^{-1} + O(\vartheta). \label{tr4}
\end{align}
\end{subequations}
Performing the limit $\vartheta \rightarrow 0$ in Eqs.
(\ref{tr3},\ref{tr4}) and combining the resulting equations with
the maps in Eqs. (\ref{tr1},\ref{tr2}) we obtain the map
(\ref{tra},\ref{trb}). Its Poisson property is ensured thanks to
the Poisson property of the map (\ref{mg43}).

The construction of the discrete Hamiltonians
(\ref{567dd},\ref{568dd}) is similar to the one done for the
continuous-time Lagrange chain. They can be obtained through the
following formulae by a straightforward computation: \bea &&
R_{i}(\ep) = \lim_{\vartheta \rightarrow 0} \,
[\,H_{2i} (\ep)+ H_{2i-1}(\ep)\,], \nonumber \\
&& S_{i}(\ep)= \lim_{\vartheta \rightarrow 0}\, [\, \vartheta \,
(\nu_{2i} \,  H_{2i}(\ep) +\nu_{2i-1} \, H_{2i-1}
(\ep))\,],\nonumber \eea $\{{H}_i(\ep) \}_{i=1}^{2M}$ being  the
Hamiltonians (\ref{disc}).

Let us finally notice that the Hamiltonians
(\ref{567dd},\ref{568dd}) are in involution w.r.t. the brackets
(\ref{xc}) thanks to Proposition \ref{7584t}.

\endpf

The discrete-time version of the Hamiltonian (\ref{H!}) is given by
\bea
\CH_{M,2}(\ep) &=& \sum_{k=1}^M [\, \, \mu_k \, R_k (\ep)+ S_k(\ep)\, ]= \nonumber\\
&=&\sum_{k=1}^M \langle \,{\bf{p}}, \mu_k \, {\bf{m}}_k + {\bf{a}}_k\, \rangle+
\half \sum_{j,k=1}^M \langle \,{\bf{m}}_j, {\bf{m}}_k \, \rangle \left(1 +
\frac{\ep^2}{4} \, \mu_j \, \mu_k \,\langle \,{\bf{p}}, {\bf{p}} \, \rangle   \right)- \nonumber\\
&& - \, \frac{\ep}{4} \sum_{\stackrel{\scriptstyle{j,k=1}} {j \neq k}}^M (\mu_k -\mu_j) \,
\, \langle \, {\bf{p}}  ,  [\,{\bf{m}}_k , {\bf{m}}_j \,]\, \rangle -
\, \frac{\ep}{2}
\, \langle \, {\bf{p}}  ,  \left[\,{\textstyle \sum_{k=1}^M} {\bf{a}}_k ,{\textstyle \sum_{j=1}^M}  {\bf{m}}_j \, \right]\, \rangle+\nonumber\\
&& + \,\frac{\ep^2}{4} \, \langle \,{\bf{p}}, {\bf{p}} \, \rangle \sum_{\stackrel{\scriptstyle{j,k=1}} {j \neq k}}^M
\mu_k  \langle \, {\bf{m}}_k , {\bf{a}}_j\, \rangle +
\, \frac{\ep^2}{8} \, \langle \,{\bf{p}}, {\bf{p}} \, \rangle\sum_{\stackrel{\scriptstyle{j,k=1}} {j \neq k}}^M
\, \langle \, {\bf{a}}_k , {\bf{a}}_j \, \rangle.
\nonumber
\eea

Notice that we still have the linear integral
$
\sum_{k=1}^M  R_k(\ep)= \sum_{k=1}^M \langle \,{\bf{p}},{\bf{m}}_k \, \rangle.
$

\section{Concluding remarks} \label{sec6}

We presented a systematic construction of finite-dimensional
integrable systems sharing the same linear $r$-matrix bracket with
the rational $\su2$ Gaudin model. The resulting one-body and
many-body integrable systems are obtained through suitable
algebraic contractions of the Lie-Poisson structure of the
ancestor model. We called these families of integrable systems
$\su2$ towers. The three-dimensional Lagrange top is  the first
element of the rational one-body $\su2$ tower. The many-body
counterpart of the Lagrange top, called Lagrange chain, is also
presented and its Lax representation is given.

In the second part of the paper we derived an explicit integrable
Poisson map discretizing a Hamiltonian flow of the rational $\su2$
Gaudin model, thus providing a new integrable discretization of
such a model. Then, the contraction procedures enable us to
construct integrable discrete-time versions of the of the rational
$\su2$ tower and of the Lagrange chain.

The main open problem connected with this work is to find Lax
representations (and then their $r$-matrix interpretation) for all
the integrable Poisson maps introduced here (actually the only
case for which the Lax representation is known is the
discrete-time Lagrange top considered in \cite{BS}). These
structures will allow to avoid a brute force verification of the
integrability, which we had to perform here. Of course, finding a
Lax representation for the discrete-time rational
$\mathfrak{su}(2)$ Gaudin model would yield the corresponding
results for all the contracted systems.

Also the following problem deserves further investigations. It is
well-known that the continuous-time rational Gaudin models, as
well as the one-body and many-body towers \cite{P} described in
the present work, admit a multi-Hamiltonian formulation \cite{FM}.
Finding a multi-Hamiltonian formulation of our discrete-time maps
is an open challenge.

\section*{Aknowledgments}

M.P. wishes to express his gratitude to F. Musso, for his constant help and support.
M.P. is also grateful to G. Satta and O. Ragnisco for  many interesting discussions.
 M.P. was partially
supported by the European Community through the FP6 Marie
Curie RTN ENIGMA (Contract number MRTN-CT-2004-5652) and by
the European Science Foundation project MISGAM.

\section*{Appendix 1: Visualization}

As shown in Proposition \ref{popore} the integrable Poisson map (\ref{mgtilde}) discretizing the rational one-body
$\su2$ tower is well defined and can be easily iterated. We present here its visualization in the case $N=2$
(i.e. the Lagrange top) and $N=3$ (i.e. the first extension of the Lagrange top).

The input parameters are:
the intensity of the external field, $p$;
the discretization parameter, $\ep$;
the number of iteration of the map, $\mathcal{N}$;
the initial values of the coordinate functions:  $({\bf{z}}_0(0),{\bf{z}}_1(0))$ for $N=2$
and $({\bf{z}}_0(0),{\bf{z}}_1(0),{\bf{z}}_2(0))$ for $N=3$.

\begin{figure}[h!]

\begin{center}

\includegraphics[height=6cm]{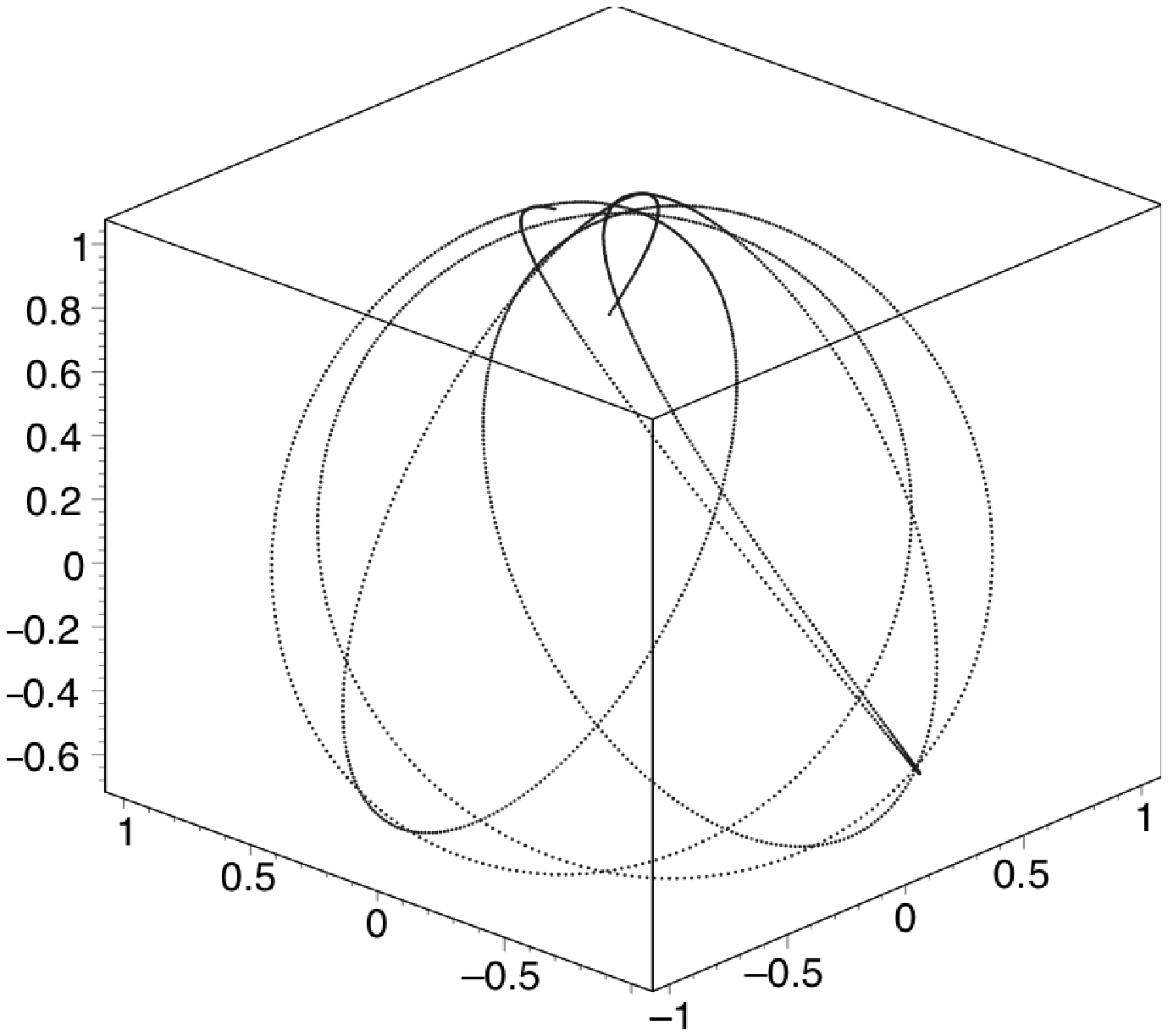}
\hspace{.5truecm}
\includegraphics[height=6cm]{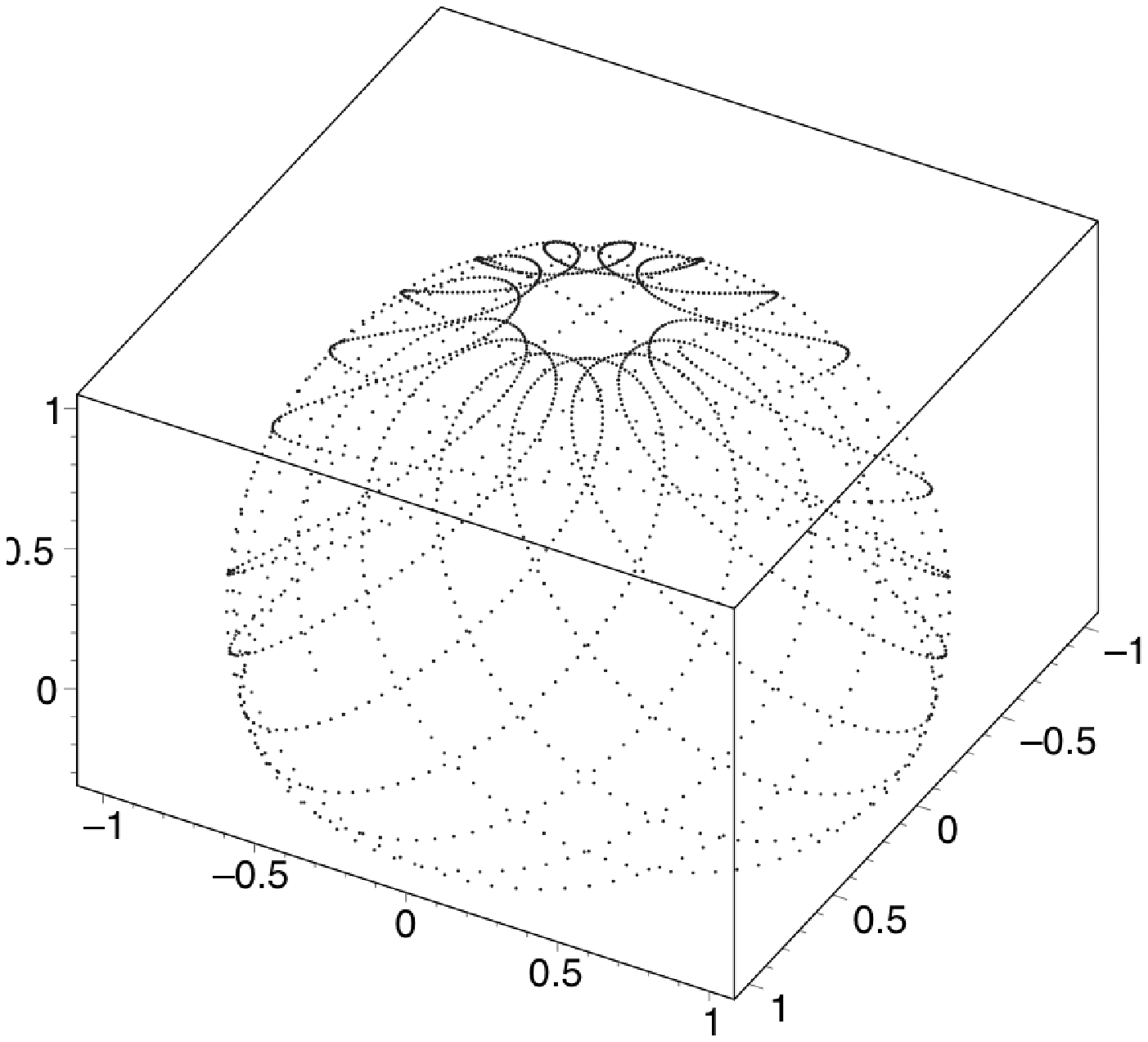}

\end{center}

\end{figure}

The first plots refer to the case $N=2$. The output is a 3D plot
of $\mathcal{N}$ consequent points $(z_1^1,z_1^2,z_1^3)$,
describing the evolution of the axis of symmetry of the top on the
surface $\langle \, {\bf{z}}_1 , {\bf{z}}_1\, \rangle$=constant.
These plots show the typical (discrete-time) precession of the
axis.

The second ones refer to the case $N=3$. The output is a 3D plot
of $\mathcal{N}$ consequent points $(z_2^1,z_2^2,z_2^3)$,
describing the evolution of the axis of symmetry of the top on the
surface $\langle \, {\bf{z}}_2 , {\bf{z}}_2\, \rangle$=constant
and $\mathcal{N}$ consequent points
$(z_1^1-z_2^1,z_1^2-z_2^2,z_1^3-z_2^3)$ describing the evolution
of the satellite.

\begin{figure*}[h!]

\begin{center}

\includegraphics[height=6cm]{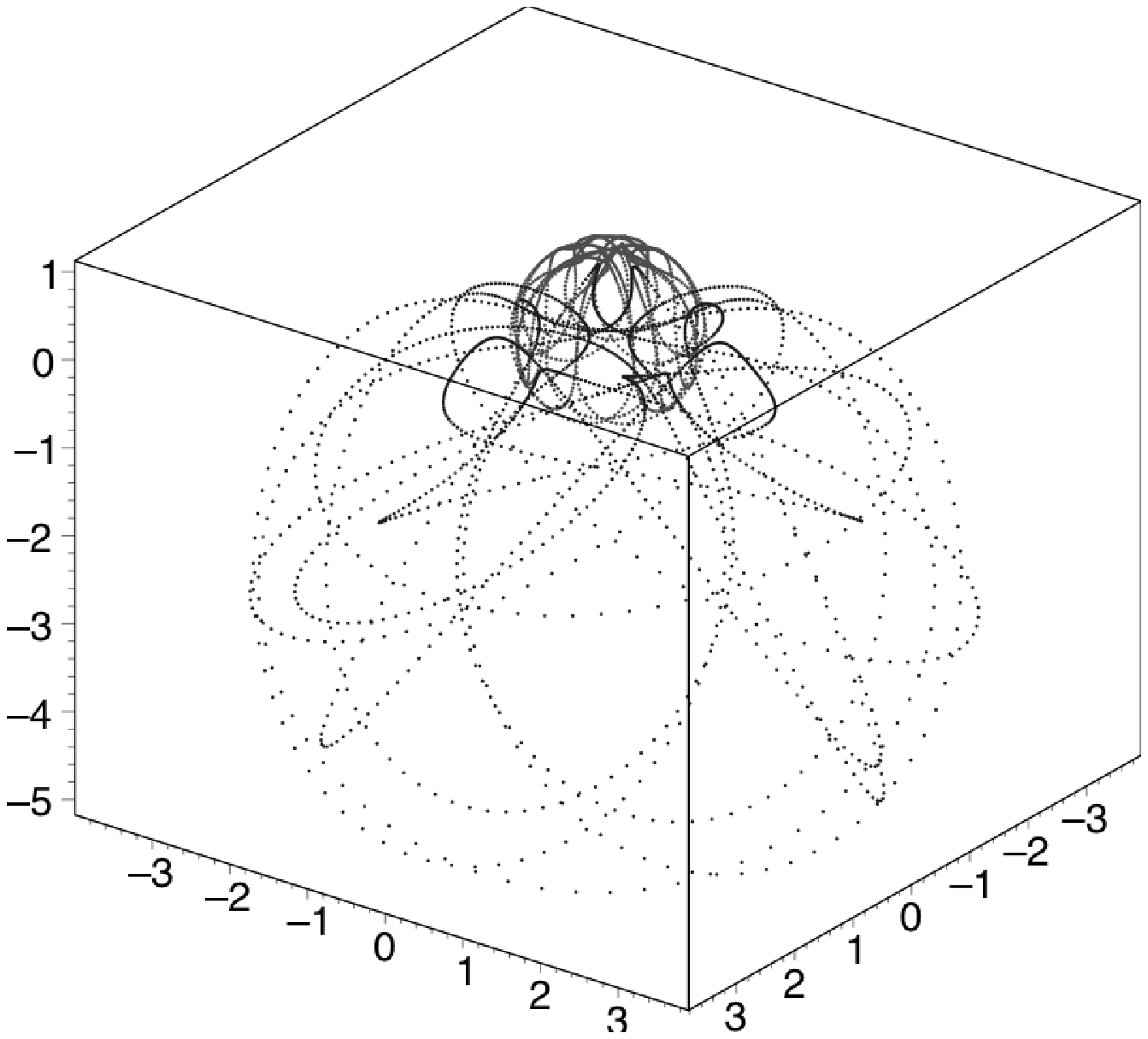}
\hspace{.5truecm}
\includegraphics[height=6cm]{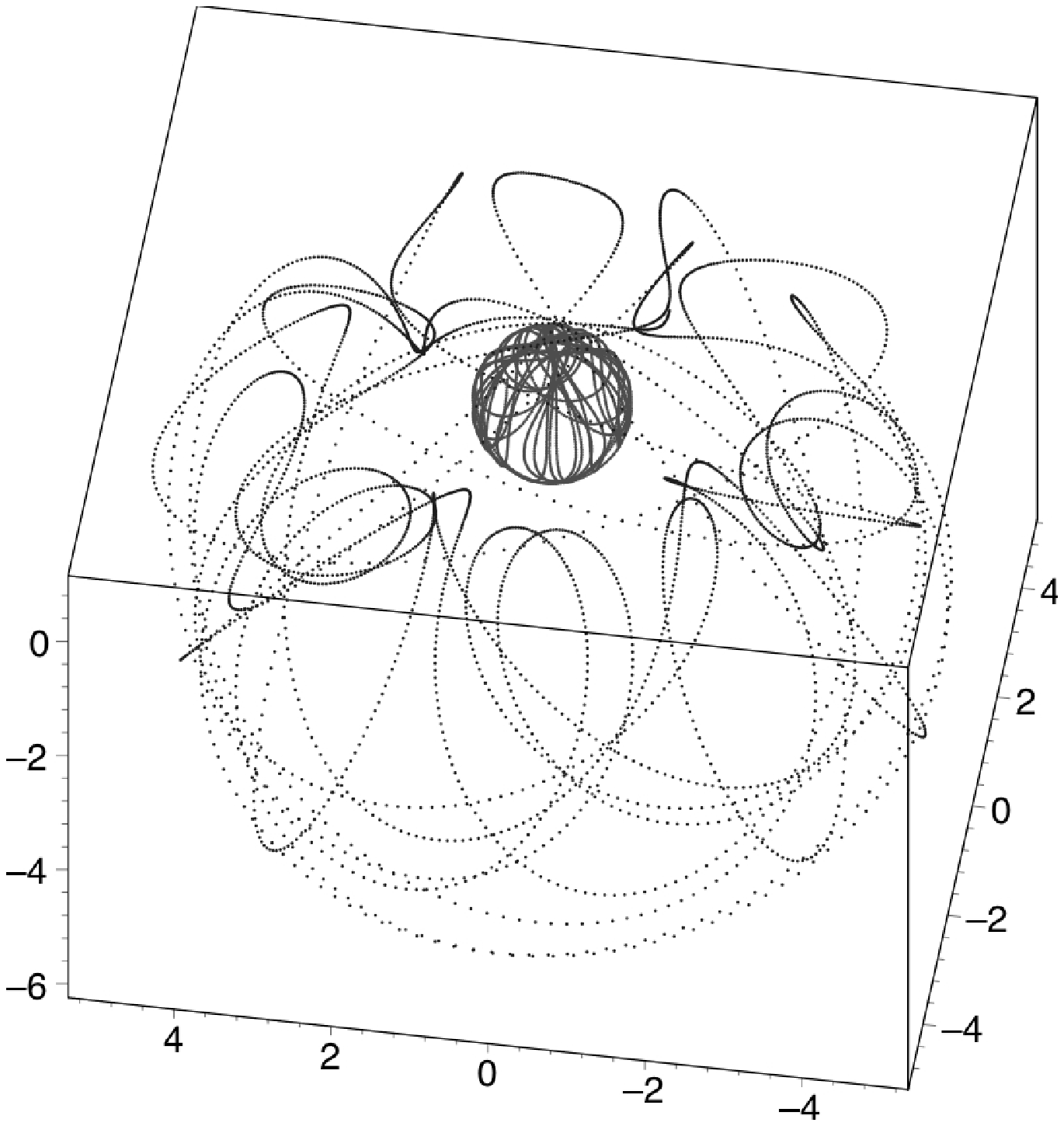}

\end{center}

\end{figure*}

Let us also give a visualization, for $M=2$, of the integrable
discrete-time evolution of the axes of symmetry of the Lagrange
tops given by the map (\ref{tra},\ref{trb}).

\begin{figure}[h!]

\begin{center}

\includegraphics[height=6cm]{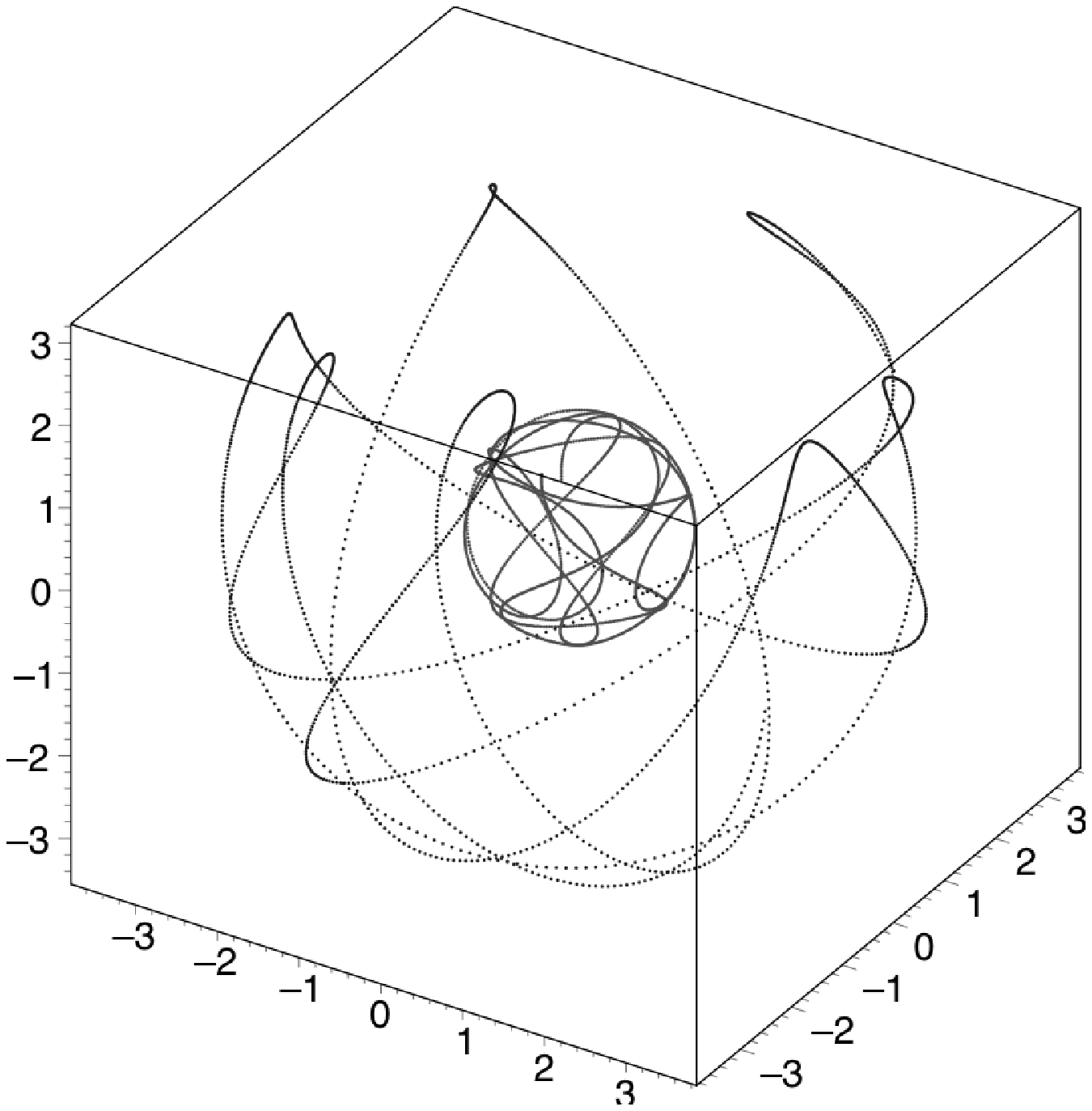}

\end{center}

\end{figure}

The input parameters are: the intensity of the external field,
$p$; the values of the parameters $\mu_1$ and $\mu_2$; the
discretization parameter, $\ep$; the number of iteration of the
map, $\mathcal{N}$; the initial values of the coordinate
functions, $({\bf{m}}_1(0),{\bf{a}}_1(0))$ and
$({\bf{m}}_2(0),{\bf{a}}_2(0))$.

The output is a 3D plot of $\mathcal{N}+\mathcal{N}$ consequent
points $(a_1^1,a_1^2,a_1^3)$ and $(a_2^1,a_2^2,a_2^3)$ describing
the evolution of the axes of symmetry of the tops respectively on
the surfaces $\langle \, {\bf{a}}_1 , {\bf{a}}_1\,
\rangle$=constant and $\langle \, {\bf{a}}_2, {\bf{a}}_2\,
\rangle$=constant.

\section*{Appendix 2: Proof of the involutivity of the functions $\{H_k(\ep)\}_{k=1}^N$}

Let us write the functions $\{H_k(\ep)\}_{k=1}^N$  given in Eq. (\ref{disc}) in the  following way:
$$
H_k(\ep) = h^{0}_k  -\frac{\ep}{2} \, h^{1}_k +
\frac{\ep^2}{4}  \, \langle \, {\bf{p}}  ,{\bf{p}} \, \rangle \,h^{2}_k, \qquad 1 \leq k \leq N,
$$
where
\begin{subequations}
\begin{align}
& h^{0}_k = \langle \,  {\bf{p}} ,  {\bf{y}}_k \, \rangle + \sum_{\stackrel{\scriptstyle{j=1}} {j \neq k}}^N
\frac{\langle \, {\bf{y}}_k,{\bf{y}}_j \, \rangle }{\l_k - \l_j}, \label{I0} \\
& h^{1}_k =\sum_{\stackrel{\scriptstyle{j=1}} {j \neq k}}^N
\, \langle \, {\bf{p}}  , [\,{\bf{y}}_k ,{\bf{y}}_j ]\, \rangle, \label{I1} \\
& h^{2}_k = \sum_{\stackrel{\scriptstyle{j=1}} {j \neq k}}^N
\frac{\l_k \, \l_j }{\l_k - \l_j} \, \langle \, {\bf{y}}_k,{\bf{y}}_j \, \rangle.\label{I2}
\end{align}
\end{subequations}
In the following computations we shall use the Lie-Poisson brackets (\ref{LPgdd}). We have:
\bea
\left\{ H_k(\ep), H_i(\ep) \right\}&=& \left\{h^{0}_k , h^{0}_i\right\}-
\frac{\ep}{2} \,\left( \left\{h^{0}_k , h^{1}_i\right\}+ \left\{h^{1}_k , h^{0}_i\right\}\right)+ \nonumber \\
&&  + \, \frac{\ep^2}{4}  \,\left[ \langle \, {\bf{p}}  ,{\bf{p}} \, \rangle \,
\left( \left\{h^{0}_k , h^{2}_i\right\}+ \left\{h^{2}_k , h^{0}_i\right\}\right) + \left\{h^{1}_k , h^{1}_i\right\} \, \right]- \nonumber \\
&& -\,  \frac{\ep^3}{8}  \, \left( \left\{h^{1}_k ,
h^{2}_i\right\}+ \left\{h^{2}_k , h^{1}_i\right\}\right)+
\frac{\ep^4}{16}  \,\langle \, {\bf{p}}  ,{\bf{p}} \, \rangle^2 \,
\left\{h^{2}_k , h^{2}_i\right\}.\label{6y7} \eea We already know
that $\left\{h^{0}_k , h^{0}_i\right\}=0$, $1 \leq k,i \leq N$,
since the integrals  $\left\{h_k^0\right\}_{k=1}^N$ are the ones
of the continuous-time $\su2$ rational Gaudin model. Let us
compute the remaining brackets in Eq. (\ref{6y7}) using Eqs.
(\ref{I0},\ref{I1},\ref{I2}) and assuming $k \neq i$. Notice that
in the brackets $\left\{h^{0}_k , h^{1}_i\right\}+ \left\{h^{1}_k
, h^{0}_i\right\}$ and $\left\{h^{0}_k , h^{2}_i\right\}+
\left\{h^{2}_k , h^{0}_i\right\}$ we shall explicitly write the
order of $|{\bf{p}}|= \langle \, {\bf{p}}  ,{\bf{p}} \,
\rangle^{1/2}=p$ appearing in the computation.

At order $\ep$ we have:
\bea
&& \; \left[\, \left\{h^{0}_k , h^{1}_i\right\}+ \left\{h^{1}_k , h^{0}_i\right\}\,
\right]_{O(| {\bf{p}}| )}= \nonumber \\
&& =
p^\beta \, \epsilon_{\beta \rho \sigma}
\sum_{\stackrel{\scriptstyle{j=1}} {j \neq k}}^N \sum_{\stackrel{\scriptstyle{l=1}}
{l \neq i}}^N
\left[\,
\frac{1}{\l_k - \l_j} \left\{ y_k^\al \, y_j^\al, y_i^\rho \, y_l^\sigma\right\} +
\frac{1}{\l_i - \l_l} \left\{ y_k^\rho \, y_j^\sigma, y_i^\al \, y_l^\al\right\}
\, \right]= \nonumber \\
&& = - p^\beta \, \epsilon_{\beta \rho \sigma} \, \epsilon_{\al \sigma \gamma}
\sum_{\stackrel{\scriptstyle{j=1}} {j \neq k}}^N
\frac{1}{\l_k - \l_j} ( y_k^\ga \, y_j^\al \, y_i^\rho + y_j^\ga \, y_i^\rho \, y_k^\al  )
- \nonumber \\
&& \quad - p^\beta \, \epsilon_{\beta \rho \sigma} \, \epsilon_{\rho \al \ga }
\sum_{\stackrel{\scriptstyle{j=1}} {j \neq k}}^N
\frac{1}{\l_i - \l_k} (y_k^\ga \, y_j^\sigma \, y_i^\al +
 y_i^\ga \, y_j^\sigma \, y_k^\al  )- \nonumber \\
&& \quad - p^\beta \, \epsilon_{\beta \rho \sigma} \, \epsilon_{\sigma \al \ga}
\sum_{\stackrel{\scriptstyle{j=1}} {j \neq k}}^N
\frac{1}{\l_i - \l_j} (y_k^\rho \, y_j^\ga \, y_i^\al +
 y_i^\ga \, y_j^\al \, y_k^\rho  ).\nonumber
\eea
The above expression vanishes if we swap the indices $\al$ and $\ga$ in each second
term in the three brackets. Then we have:
\bea
\left[\,\left\{h^{0}_k , h^{1}_i\right\}+ \left\{h^{1}_k , h^{0}_i\right\}\,
\right]_{O(| {\bf{p}}|^2 )}&=&
p^\al \, p^\beta \, \epsilon_{\beta \rho \sigma} \sum_{l=1}^N
\left[\,
\left\{ y_k^\al, y_i^\rho \, y_l^\sigma\right\} +
\left\{ y_k^\rho \, y_l^\sigma, y_i^\al\right\}
\, \right] = \nonumber \\
&=& p^\al \, p^\beta \, (\epsilon_{\beta \rho \sigma} \,
\epsilon_{\al \sigma \ga} + \epsilon_{\beta \ga \sigma} \,
\epsilon_{\sigma \al \rho})\, y_k^\ga \, y_i^\rho, \nonumber \eea
that vanishes due to the properties of the tensor $\epsilon_{\al
\be \ga }$.

At order $\ep^2$ we get:
\bea
 && \; \left[\,\left\{h^{0}_k , h^{2}_i\right\}+ \left\{h^{2}_k , h^{0}_i\right\} \,
 \right]_{O(| {\bf{p}}| )}= \nonumber \\
&& = p^\al
\sum_{\stackrel{\scriptstyle{l=1}} {l \neq i}}^N \frac{\l_i \, \l_l}{\l_i-\l_l}
\left\{ y_k^\al, y_i^\be y_l^\be\right\}
-p^\al
\sum_{\stackrel{\scriptstyle{j=1}} {j \neq k}}^N \frac{\l_k \, \l_j}{\l_k-\l_j}
\left\{ y_i^\al, y_k^\be y_j^\be\right\}= \nonumber \\
&& = -p^\al \epsilon_{\al \be \ga} \left( \frac{\l_i \, \l_k}{\l_i-\l_k}y_i^\be y_k^\ga  +
\frac{\l_i \, \l_k}{\l_i-\l_k}y_i^\ga y_k^\be  \right), \nonumber
\eea
that vanishes swapping the indices $\gamma$ and $\beta$ in the second term. Moreover,
\bea
&& \; \left[\,\left\{h^{0}_k , h^{2}_i\right\}+ \left\{h^{2}_k , h^{0}_i\right\} \,
\right]_{O(| {\bf{p}}|^0 )}=
 \sum_{\stackrel{\scriptstyle{j=1}} {j \neq k}}^N \sum_{\stackrel{\scriptstyle{l=1}}
 {l \neq i}}^N
\frac{\l_i \, \l_l + \l_k \, \l_j }{(\l_k - \l_j)(\l_i-\l_l)}
\left\{ y_k^\al \, y_j^\al, y_i^\be \, y_l^\be\right\} =\nonumber \\
&& = - \epsilon_{\al \be \ga} \sum_{\stackrel{\scriptstyle{j=1}} {j \neq k}}^N
\frac{\l_k (\l_i^2 -\l_j^2) -\l_i(\l_k^2 -\l_j^2) -\l_j(\l_i^2-\l_k^2)}
{(\l_k - \l_j)(\l_i-\l_k)(\l_i-\l_j)}
y_k^\ga \, y_j^\al \,  y_i^\be =\nonumber \\
&& = - \epsilon_{\al \be \ga} \sum_{\stackrel{\scriptstyle{j=1}} {j \neq k}}^N y_k^\ga \,
y_j^\al \,  y_i^\be. \nonumber
\eea
On the other hand:
\bea
&& \left\{h^{1}_k , h^{1}_i\right\}= p^\al \, p^\sigma \,
\epsilon_{\al \be \ga}\epsilon \, _{\sigma \rho \mu}
\sum_{\stackrel{\scriptstyle{j=1}} {j \neq k}}^N
\sum_{\stackrel{\scriptstyle{l=1}} {l \neq i}}^N
\left\{ y_k^\al\, y_j^\ga, y_i^\rho \, y_l^\mu\right\} =
p^\sigma \, p^ \sigma\,
 \epsilon_{\al \be \ga} \sum_{\stackrel{\scriptstyle{j=1}}
 {j \neq k}}^N y_k^\ga \, y_j^\al \,  y_i^\be,\nonumber
\eea
where we have used the properties of the tensor $\epsilon_{\al \be \ga }$. Hence we get:
$$
\langle \, {\bf{p}}  ,{\bf{p}} \, \rangle \,
\left( \left\{h^{0}_k , h^{2}_i\right\}+ \left\{h^{2}_k , h^{0}_i\right\}\right)
+ \left\{h^{1}_k , h^{1}_i\right\}=0.
$$

At order $\ep^3$ we have:
\bea
&& \left\{h^{1}_k , h^{2}_i\right\}+ \left\{h^{2}_k , h^{1}_i\right\}= \nonumber \\
&& =
-p^\beta \, \epsilon_{\beta \rho \sigma}
\sum_{\stackrel{\scriptstyle{j=1}} {j \neq k}}^N \sum_{\stackrel{\scriptstyle{l=1}}
{l \neq i}}^N
\left[
\frac{\l_k \, \l_j }{\l_k - \l_j} \left\{ y_k^\al \, y_j^\al, y_i^\rho \, y_l^\sigma\right\} +
\frac{\l_i \, \l_l}{\l_i - \l_l} \left\{ y_k^\rho \, y_j^\sigma, y_i^\al \, y_l^\al\right\}
\, \right]= \nonumber \\
&& = p^\beta \, \epsilon_{\beta \rho \sigma} \, \epsilon_{\al \sigma \gamma}
\sum_{\stackrel{\scriptstyle{j=1}} {j \neq k}}^N
\frac{\l_k \, \l_j}{\l_k - \l_j}
( y_k^\ga \, y_j^\al \, y_i^\rho + y_j^\ga \, y_i^\rho \, y_k^\al  ) - \nonumber \\
&& \quad -\,p^\beta \, \epsilon_{\beta \rho \sigma} \, \epsilon_{\rho \al \ga }
\sum_{\stackrel{\scriptstyle{j=1}} {j \neq k}}^N
\frac{ \l_k \, \l_i}{\l_i - \l_k} (y_k^\ga \, y_j^\sigma \, y_i^\al +
 y_i^\ga \, y_j^\sigma \, y_k^\al  )- \nonumber \\
&& \quad -\, p^\beta \, \epsilon_{\beta \rho \sigma} \, \epsilon_{\sigma \al \ga}
\sum_{\stackrel{\scriptstyle{j=1}} {j \neq k}}^N
\frac{\l_i \, \l_j}{\l_i - \l_j} (y_k^\rho \, y_j^\ga \, y_i^\al +
 y_i^\ga \, y_j^\al \, y_k^\rho  ).\nonumber
\eea
The above expession vanishes if we swap the indices $\al$ and $\ga$ in each second
term in the three brackets.

Finally, at order $\ep^4$, we get:
\bea
&& \left\{h^{2}_k , h^{2}_i\right\}=
\sum_{\stackrel{\scriptstyle{j=1}} {j \neq k}}^N \sum_{\stackrel{\scriptstyle{l=1}}
{l \neq i}}^N
\frac{\l_k \, \l_j \, \l_i \, \l_l}{(\l_k - \l_j)(\l_i - \l_l)}
\left\{ y_k^\al \, y_j^\al, y_i^\be \, y_l^\be\right\}= \nonumber \\
&&= -\epsilon_{\al \be \ga}\sum_{\stackrel{\scriptstyle{j=1}} {j
\neq k}}^N y_j^\al \, y_i^\be \, y_k^\ga \left[\, \frac{\l_k^2 \,
\l_j \, \l_i}{(\l_i - \l_k)(\l_k - \l_j)} + \frac{\l_k \, \l_j \,
\l_i^2}{(\l_i - \l_j)(\l_k - \l_i)} - \frac{\l_k\, \l_j^2 \,
\l_i}{(\l_i - \l_j)(\l_k - \l_j)} \, \right].\nonumber \eea A
direct computation shows that the expression in the square
brackets vanishes.



\begin{thebibliography}{99}

\bibitem{AHH}
Adams M., Harnad J., Hurtubise J.,
Comm. Math. Phys. {\bf 155} (1993) 385--413.

\bibitem{AO}
Amico L., Osterloh A.,
Phys. Rev. Lett. {\bf 88} (2002) 127003.

\bibitem{A}
Audin M.,
Spinning tops, Cambridge University Press, 1996.

\bibitem{BMR}
Ballesteros A., Ragnisco O.,
Jour. Phys. A {\bf 31} (1998) 3791--3813.

Ballesteros A., Musso F., Ragnisco O.,
Jour. Phys. A {\bf 39} (2002) 8197--8211.


\bibitem{BD}
Belavin A.A., Drinfel'd V.G.,
Funktsional. Anal. i Prilozhen {\bf 16} (1982) 1--29.

\bibitem{BS}
Bobenko A.I., Suris Yu.B.,
Comm. Math. Phys. {\bf 204} (1999) 147--188.

\bibitem{BMc}
Brzezinzki T., Macfarlane A.J.,
Jour. Math. Phys. {\bf 35} (1994) 3261--3275.


\bibitem{Ch}
Chernyakov Yu.B.,
Theor. Math. Phys. {\bf 141} (2004) 1361--1380.

\bibitem{EFR}
Enriquez B., Feigin B., Rubtsov V.
Comp. Math. {\bf 110} (1998) 1--16.

\bibitem{FM}
Falqui G., Musso F.,
Jour. Phys. A {\bf 36} (2003) 11655--11676.

\bibitem{FFR}
Feigin F., Frenkel E., Reshetikhin N.Yu.,
Comm. Math. Phys. {\bf 166} (1994) 27--62.

\bibitem{Fr}
Frenkel E., in
Proceedings of ``XIth international congress of mathematical physics'', International Press,
1995, 606--642.

\bibitem{KMa}
Kulish P., Manojlovic N.,
Jour. Math. Phys. {\bf 42} (2001) 4757--4774.

\bibitem{G1}
Gaudin M.,
Jour. de Phys. {\bf 37} (1976) 1087--1098.

\bibitem{G2}
Gaudin M.,
La fonction d' onde de Bethe, Masson, Paris, 1983.


\bibitem{GZ}
Gavrilov L., Zhivkov A.,
L' Enseign. Math. {\bf{44}}  (1998) 133--170.

\bibitem{Ge}
Gekhtman M.I.,
Comm. Math. Phys. {\bf 167} (1995) 593--605.

\bibitem{HKR}
Hone A.N.W., Kuznetsov V.B., Ragnisco O.,
Jour. Phys. A {\bf 34} (2001) 2477--2490.

\bibitem {IW}
In\"on\"u E., Wigner E.P.,
Proc. Natl Acad. Sci. {\bf 39} (1953) 510--24.

\bibitem{J}
Jur\v{c}o B.,
Jour. Math. Phys. {\bf 30} (1989) 1289--1293.

\bibitem{KKM2}
Kalnins E.G., Kuznetsov V.B., Miller W.Jr.,
Jour. Math. Phys. {\bf 35} (1994) 1710--1731.

\bibitem{KM}
Kapovich M., Milson J.,
Journ. Diff. Geom. {\bf 44} (1996) 479--513.

\bibitem{KPR}
Kuznetsov V.B., Petrera M., Ragnisco O.,
Jour. Phys. A {\bf 37} (2004) 8495--8512.

\bibitem{KS1}
Kuznetsov V.B., Sklyanin E.K.,
Jour. Phys. A {\bf 31} (1998) 2241--2251.


\bibitem{MV}
Moser J., Veselov A.P.,
Comm. Math. Phys. {\bf 139} (1991)  217--243.

\bibitem{MPR}
Musso F., Petrera M., Ragnisco O.,
Jour. Nonlinear Math. Phys. {\bf 12} suppl. 1 (2005) 482--498.

\bibitem{MPRS2}
Musso F., Petrera M., Ragnisco O., Satta G.,
Regul. Chaotic Dyn. {\bf{10}} 4 (2005) 363--380.

\bibitem{MPRS3}
Musso F., Petrera M., Ragnisco O., Satta G.,
Jour. Nonlinear Math. Phys. {\bf 12} suppl. 2 (2005) 240--252.

\bibitem{MPRSq}
Musso F., Petrera M., Ragnisco O., Satta G.,
Nucl. Phys. B {\bf 716} (2005) 543--555.

\bibitem{Ne}
Nekrasov N.,
Comm. Math. Phys. {\bf 180} (1996) 587--603.

\bibitem{P}
Petrera M.,
Integrable extensions and discretizations of classical Gaudin models,
PhD Thesis, 2007, Physics Department, University of Roma III.

\bibitem{PR}
Petrera M., Ragnisco O.,
Sigma {\bf 3} (2007) 058 14 pages.

\bibitem{RSD}
Roman J.M., Sierra G., Dukelski J.,
Nucl. Phys. B  {\bf 634} (2002) 483--510.

\bibitem{RSTS}
Reyman A.G., Semenov-Tian-Shansky M.A.,
Group theoretical methods in the theory of finite-dimensional integrable systems, in
Dynamical systems VII, Springer, 1994.

\bibitem{Re}
Reshetikhin N.Yu.,
Lett. Math. Phys. {\bf 26} (1992) 167--177.

\bibitem{SW}
Seiberg N., Witten E.,
Nucl. Phys. B {\bf{426}}  (1994) 19--35.

\bibitem{S0}
Sklyanin E.K.,
Jour. Sov. Math. {\bf{19}} (1982) 1546--1596.

\bibitem{S1}
Sklyanin E.K.,
Jour. Sov. Math. {\bf{47}} (1989)  2473--2488.

\bibitem{ST}
Sklyanin E.K., Takebe T.,
Phys. Lett. A {\bf{219}}  (1996) 217--225.

\bibitem{Su}
Suris Yu.B.,
The problem of integrable discretization: Hamiltonian approach,
Progress in Mathematics, {\bf{219}}, Birkh\"auser Verlag, Basel, 2003.

\bibitem{Ta}
Talalaev D.,
Preprint, www.arxiv.org/hep-th/0404153.

Chervov A., Talalaev D.,
Preprint, www.arxiv.org/hep-th/0409007.

Chervov A., Rybnikov L., Talalaev D.,
Preprint, www.arxiv.org/hep-th/0404106.

\bibitem{V1}
Veselov A.P.,
Uspekhi Mat. Nauk. {\bf 46} (1991) 3--45.

Veselov A.P.,
Comm. Math. Phys. {\bf 145} (1992) 181--193.

Veselov A.P.,
Funct. Anal. Appl.  {\bf 22} (1998) 1--13.

\bibitem{eve}
Weimar-Woods E.,
Journ. Math. Phys. {\bf 36} 8 (1995) 4519--4548.


\end{thebibliography}
\end{document}